\documentclass[aps,pra,reprint,onecolumn,superscriptaddress,nofootinbib]{revtex4-2}
\usepackage{amsmath}
\usepackage{mathtools}
\usepackage{graphicx}
\usepackage{amsfonts}
\usepackage{color}
\usepackage{microtype}
\usepackage{array}
\usepackage{tabularx}
\usepackage{booktabs}
\usepackage{cryptocode}
\usepackage{enumitem}
\usepackage{verbatim}

\usepackage{xparse}
\usepackage{xspace}
\usepackage{braket}
\usepackage{physics}
\usepackage{bm}

\usepackage{mdframed}

\makeindex
\usepackage{amsthm}
\usepackage{thmtools}
\usepackage{thm-restate}
\usepackage{tikz}
\usetikzlibrary{positioning, calc, shapes.geometric, backgrounds, fit, arrows.meta}

\usepackage{titlesec}

\makeatletter
\renewcommand{\p@subsection}{}
\renewcommand{\p@subsubsection}{}
\makeatother
\titleformat*{\paragraph}{\bfseries}

\renewcommand{\selectlanguage}[1]{} 

\usepackage{hyperref} 
\usepackage[capitalize]{cleveref} 

\crefname{enumi}{}{}
\hypersetup{
	colorlinks = true,
	linkcolor =blue,
	citecolor=blue,
	urlcolor=blue
}

\newlength\colsep   
\newlength\offsetX  
\newlength\offsetY  

\newtheorem{theorem}{Theorem}[section]
\newtheorem{lemma}{Lemma}[section]
\newtheorem{corollary}{Corollary}[section]
\newtheorem{remark}{Remark}[section]

\newtheorem{condition}{Condition}
\crefname{condition}{Condition}{Conditions}
\theoremstyle{definition} 
\newtheorem{definition}{Definition}[section]

\newcommand{\Fobs}{F_{\mathrm{obs}}}
\newcommand{\Fobstest}{F_{\mathrm{obs}}^{\mathrm{test}}}
\newcommand{\Fobsgen}{F_{\mathrm{obs}}^{\mathrm{gen}}}
\newcommand{\Ftest}{F_0^{\mathrm{test}}}
\newcommand{\Fgen}{F_0^{\mathrm{gen}}}

\newcommand{\floor}[1]{\left\lfloor #1 \right\rfloor} 
\newcommand{\ceil}[1]{\left\lceil #1 \right\rceil} 
\newcommand{\Renyi}{R\'{e}nyi}
\NewDocumentCommand{\Halpha}{o d<>}{%
  \ensuremath{%
    \widetilde{H}^{\uparrow\IfValueT{#1}{\,,#1}}_{\IfValueTF{#2}{#2}{\alpha}}%
  }%
}
\NewDocumentCommand{\HalphaDown}{o}{%
  \ensuremath{\widetilde{H}^{\downarrow\IfValueT{#1}{\,,#1}}_\alpha}}

\newcommand{\dop}[1]{\operatorname{S}_{#1}}

\newcommand{\CPTP}{\operatorname{CPTP}}
\newcommand{\id}{\operatorname{id}}
\newcommand{\defvar}{\coloneqq}

\newcommand{\pf}{\operatorname{\mathtt{Pur}}}

\newcommand{\term}[1]{\textbf{#1}}

\newcommand{\mbf}[1]{\mathbf{#1}}

\newcommand{\sourcesymbol}{\sigma}

\newcommand{\CP}{\widehat{C}}
\newcommand{\CPhat}{\widetilde{C}}
\newcommand{\Ameas}{\bar{A}}
\newcommand{\Ashield}{\hat{A}}
\newcommand{\alphCP}{\widehat{\mathcal{C}}}
\newcommand{\cP}{\hat{c}}

\newcommand{\attack}[1]{\mathcal{A}_{#1}}
\newcommand{\QKDGmap}[1]{\mathcal{T}_{#1}}
\newcommand{\onerndchann}{\mathcal{M}}

\newcommand{\fhat}{\hat{f}}
\newcommand{\fhatfull}{\fhat_\mathrm{full}}

\newcommand{\sigmaconstraint}[1]{\sourcesymbol_{A_{#1}}^{(#1)}}

\newcommand{\alphafinal}{\alpha_\mathrm{f}}
\newcommand{\Mem}{M}

\newcommand{\EffMem}{ \widetilde{M}}
\newcommand{\effmem}{\widetilde{m}}
\newcommand{\effmemSet}{\widetilde{\mathcal{M}}}
\newcommand{\coarsegrainedhistory}[1]{\mathcal{T}^\mathrm{coarse}_{#1}}
\newcommand{\createhistory}[1]{\mathcal{T}^\mathrm{nextM}_{#1}}

\newcommand{\historyupdatemap}[1]{u_{#1}}

\newcommand{\effectiveQKDGmap}[1]{\mathcal{T}_{#1}^{\mathrm{eff}}}

\newcommand{\projmap}{\Pi}

\NewDocumentCommand{\kappafuncgeneric}{m m m m O{\mathrm{\alpha}}}{\kappa_{#5}\!\left(#1,\;#2,\;#3,\;#4\right)
}

\newcommand{\corrperiod}{L}
\newcommand{\interleavedrounds}[1]{\mathcal{I}_{#1}}
\newcommand{\otherinterleavedrounds}[1]{\mathcal{I}_{\neq #1}}
\newcommand{\selectedround}[2]{\widetilde{j}_{#1,#2}}
\newcommand{\sourceblock}[2]{\mathcal{B}_{#1,#2}}
\newcommand{\superchannel}[2]{\mathcal{N}_{#1,#2}}
\newcommand{\blockversion}[1]{\mathbf{#1}}

\newcommand{\measmap}[1]{\mathcal{P}^{\mathrm{meas}}_{#1}}

\definecolor{llmrevision}{rgb}{0.45,0.12,0.60}

\newcommand{\measuredsource}[1]{\sourcesymbol^{\mathrm{meas},#1}}

\newcommand{\jlast}{j_\mathrm{end}}
\newcommand{\jstart}{j}
\newcommand{\lcdet}{l^\mathrm{det}_\mathrm{c}}
\newcommand{\lcsrc}{l^\mathrm{src}_\mathrm{c}}
\newcommand{\discard}{\bot^\mathrm{discard}}

\newcommand{\stopannouncement}{\bot^\mathrm{stop}}

\begin{document}

\title{MEAT with source and detector correlations}
\author{Devashish Tupkary}
\email{devashish.jayant.tupkary@umontreal.ca}
	 \affiliation{Institute for Quantum Computing and Department of Physics and Astronomy, University of Waterloo, Waterloo, Ontario, Canada, N2L 3G1}	
     \affiliation{Department of Computer Science and Operations Research,
Université de Montréal,
Montréal, Quebec, Canada, H3T 1J4}

 \author{Amir Arqand}
	 \email{aarqand@uwaterloo.ca}
	 \affiliation{Institute for Quantum Computing and Department of Physics and Astronomy, University of Waterloo, Waterloo, Ontario, Canada, N2L 3G1}


\begin{abstract}
We prove variants of the marginal-constrained entropy accumulation
theorem (MEAT) that enable
finite-size security analysis of quantum key distribution with source and detector correlations.
For source correlations, our results allow for
the state prepared in a given round to depend on the current
setting choice and a finite number of preceding choices. For detector correlations, our results require memory effects to persist for a finite number of rounds after a registered click, and 
all detectors to be switched off during those rounds. These results can be combined, providing, to the best of our knowledge, the first method to jointly treat correlated source and detector imperfections in quantum key distribution. Moreover, we retain all the benefits of the MEAT, such as (but not limited to) support for on-the-fly announcements and choice of adaptive tradeoff functions. The resulting key rates can be computed via a single-round optimization amenable to existing numerical methods. We demonstrate these results on qubit BB84, and exhibit competitive key rates even in the presence of correlated source and detector imperfections. 
\end{abstract}

\maketitle

\section{Introduction}
Most quantum key distribution (QKD) security analyses are applicable to protocols where Alice and Bob perform \emph{independent} operations in each round. In such protocols, in each round, Alice samples independently from a probability distribution, prepares the appropriate signal state, and transmits it to Bob. In each round Bob performs an independent measurement and records his outcome. Alice and Bob then communicate classically and perform local classical operations to generate keys that equal to one another, and secret from Eve (or abort). 

There are several known effects that result in Alice's and Bob's actions in each round being correlated to previous rounds. For instance, Bob uses threshold detectors to measuring incoming photons. These threshold detectors exhibit correlated effects such as dead times and afterpulsing \cite{scarani_security_2008,zapatero2023implementationsecurityquantumkey,xu_secureqkd_2020}. Dead times refer to the time window during which a detector is unresponsive after registering a click; if the next round is measured within this window, then the measurement performed in this round depends on the outcome of the previous measurement. Similarly, afterpulsing refers to the effect where a detection event increased the probability of dark counts (spurious clicks when no photons are present)
for some time. In such cases, Bob's measurement in a given round is not independent of previous rounds.
On the side of the source, Alice's prepared state in round $j$ may not be perfectly independent of her actions in the previous round; for instance, it may depend on the polarization or intensity setting of the previous rounds.

While there is a large amount of literature on the important problem of imperfections in QKD\cite{tupkary_phase_2024,CurrasLorenzo_securityframeworkimperfect_2025,wang2025phase,zapatero2023implementationsecurityquantumkey,nahar_phd_2025,kamin_renyi_2025,navarrete2026numericalsecurityanalysispractical,pereira2025optimalkeyratesquantum}, most of it is restricted to independent imperfections. There is comparatively little work on correlated imperfections \cite{curraslorenzo_phaseerrorcorrelated_2026,wang2025phase,marwah2024proving} (which we discuss in greater detail below). 
Thus, extending proof techniques to the correlated imperfections scenario is an important open task. 
In this work, we prove two variations of the MEAT
\cite{arqand_marginal_2025}
(\cref{theorem:meatwithsourcecorrelations,thm:meat-effective-history})
that enable the analysis of QKD protocols with source and
detector correlations. Our proofs work by reducing the new settings to applications of the original MEAT statement. As applied to QKD, both
results strengthen the MEAT statement and retain its other benefits,
including the ability to incorporate generic on-the-fly announcements and 
adaptive choice of tradeoff functions \cite{BDR3,tupkary_securityproofs_2026}, resilience to independent
imperfections  and
tight performance \cite{kamin_renyi_2025,wiesemann_securityframework_2026}, and modularity. Thus, they represent some of the most advanced results on finite-size QKD security with correlated imperfections to date (see \cref{subsec:relatedwork} below).

This work is structured as follows. In \cref{subsec:relatedwork} below, we
discuss prior work on these topics. In \cref{sec:background}, we
introduce notation, state the information-theoretic tools we use in this work, and
explain how the MEAT is used for QKD security analysis. In \cref{sec:detector-memory}, we
analyze detector correlations, prove
\cref{thm:meat-effective-history}, and explain how to apply it 
to QKD protocols. We also prove an improved  version in the form of 
\cref{thm:meat-effective-history-nomarginal}, which is suitable for
entanglement-based protocols. A direct application of our results requires
a finite correlation length and detectors to be shut off
following a registered click. The resulting setting is similar to that
of Ref.~\cite{wang2025phase}. In \cref{sec:source-correlations}, we
study source correlations, prove
\cref{theorem:meatwithsourcecorrelations}, and explain how to apply it to QKD protocols. We
consider the same setting as Ref.~\cite{curraslorenzo_phaseerrorcorrelated_2026}, where the signal state prepared in a given round can depend on the history of previous setting choices, for some finite correlation length. Our overall approach is also inspired by that work. In \cref{sec:combined}, we
combine the two results and prove \cref{theorem:combinedstatement}. In \cref{sec:plots}, we apply these results to obtain key rates for the qubit BB84 protocol in the presence of correlated source and detector imperfections. The resulting rates demonstrate that useful key rates remain achievable under these imperfections. To the best of our knowledge, this is the first analysis to jointly account for these imperfections.
We conclude in \cref{sec:conclusion}.

In this work, we focus our attention on the somewhat abstract statements and the assumptions under which they hold. We explain how to use the theorems in a QKD security analysis and illustrate their application through key rate calculations for qubit BB84 with correlated source and detector imperfections. We leave detailed studies of specific hardware models, including the characterization of imperfections and the choice of experimentally motivated parameters, to future work \cite{inprep_correlatedjerome}. To facilitate these studies, we largely follow the notation of Refs.~\cite{BDR3,wiesemann_securityframework_2026}, and utilize their results on QKD security and independent imperfections.

\subsection{Related Work} \label{subsec:relatedwork}

Here, we briefly review prior approaches to handling correlated imperfections (see also \cite[Section 7]{tupkary_securityproofs_2026}). The postselection technique \cite{christandl_postselection_2009,nahar_postselection_2024} requires IID measurements to ensure a crucial permutation invariance property of the protocol. It therefore appears to be unable to handle correlated imperfections without a reduction to the IID measurement setting. 
Existing entropy accumulation theorems \cite{dupuis_entropy_2020,arqand_generalized_2024,arqand_marginal_2025,metger_generalised_2022} cannot directly accommodate the correlated scenarios described above. However, Ref.~\cite{marwah2024proving} establishes a result for qubit sources with arbitrary correlations. Its approach requires hardware modifications and yields low key rates; extending it to practical sources and improving its performance remain open problems. Ref.~\cite{nahar2025imperfect} only sketches an approach to handling detector correlations using the MEAT, which we follow up on and refine in this work. Proof techniques based on phase error correction
\cite{koashi_simple_2009} or entropic uncertainty relations
\cite{tomamichel_uncertainty_2011} require a suitable bound on the so called phase
error rate. These techniques have recently been extended to several
correlated scenarios
\cite{curraslorenzo_phaseerrorcorrelated_2026,pereira2020quantum,curraslorenzo2026securitydecoystatequantumkey}.
In particular, Ref.~\cite{curraslorenzo_phaseerrorcorrelated_2026}
reduces the problem of bounding the phase error rate for a source with
finite correlation length to the corresponding problem for independent
source imperfections in multiple interleaved protocols. Our treatment
of source correlations in \cref{sec:source-correlations} is inspired by
this approach. The cited analyses address correlations associated only with
the bit and basis choices, and cannot accommodate
correlations associated with the intensity choice for phase-randomized lases pulses. For detector correlations with finite correlation length, Ref.~\cite{wang2025phase} analyzes a setting in which detectors are temporarily disabled and rounds occurring too soon after a previous detection are discarded to avoid the effects of correlated detector behavior. Overall, the literature on phase-error correction typically does not tolerate on-the-fly announcements, especially in the presence of imperfections, with Ref.~\cite{wang2025phase} allowing for a restricted type of on-the-fly announcements only (with imperfections), and Ref.~\cite{tamaki_security_2018,mizutani2025protocolleveldescriptionselfcontainedsecurity} restricted to ideal hardware.

In comparison, this work can handle arbitrary on-the-fly announcements, covers the correlated detector scenario considered in Ref.~\cite{wang2025phase}, and covers and extends the correlated source scenario considered in Ref.~\cite{curraslorenzo_phaseerrorcorrelated_2026} to mixed state preparation and correlations dependent on \textit{any} setting choices. Moreover, we can combine our treatment of source and detector correlations, which, to the best of our knowledge, has not been demonstrated using any technique before. However, our results still depend on several assumptions about the correlated model we assume, as we explain in detail throughout this work. Relaxing these model assumptions to more scenarios of correlated behaviour remains an important open challenge.


\section{Notation and Background} \label{sec:background}
\subsection{Basic Notation}
We typically use uppercase letters such as $X, A, B$ to denote registers. For a sequence of registers, we use subscripts to indicate indices: $X_j$ denotes the $j$-th register, while $X_i^j$ denotes the collection of registers $X_i, X_{i+1}, \ldots, X_j$.  We use $S_=(A)$ to denote the set of density operators on the register $A$, and $S_{\leq}(A)$ to denote the set of subnormalized density operators on the register $A$. Subscripts specify the registers on which the state exists; for example, $\rho_A$ denotes a state on register $A$. For a multipartite state $\rho_{ABC\cdots}$, we write $\rho_A$ to denote its marginal on register $A$. Conversely, for a state $\rho_A$, we use $\rho_{AB}$ to denote some extension of $\rho_A$ to an additional register $B$.   We denote by $\CPTP(A,B)$ the set of CPTP maps from registers $A$ to $B$. We use $[n]$ to denote the set of natural numbers up to $n$.  

For a state $\rho \in \dop{\leq}(CQ)$ classical on $C$, written in the form
$\rho_{CQ} = \sum_c \lambda_c \ketbra{c}{c} \otimes \sigma_c$ 
for some $\sigma_c \in \dop{=}(Q)$ and $\lambda_c \geq 0$,
and an event $\Omega$ defined on the register $C$, we will define a corresponding \term{conditional state} as,
\begin{align}
\rho_{|\Omega} \defvar \frac{\Tr[\rho]}{\Tr[\rho_{\land\Omega}]} \rho_{\land\Omega} = \frac{
\sum_{c} \Tr[ \lambda_c \sigma_c]
}{\sum_{c\in\Omega} \Tr[ \lambda_c \sigma_c]} \rho_{\land\Omega}, \qquad\qquad  \text{ where }
\rho_{\land\Omega} \defvar \sum_{c\in\Omega} \lambda_c \ketbra{c}{c} \otimes \sigma_c.
\end{align}
We use $\Omega(x)$ to denote the event that a classical register $X$ stores the value $x$.
\subsection{Basic Information Theory}
We start by defining the  $f$-weighted  sandwiched {\Renyi} conditional entropies.
\begin{definition}($f$-weighted {\Renyi} entropies \cite[Definition~4.1]{arqand_marginal_2025} )
\label{def:fweighted}
Let $\rho \in \dop{=}(\CP Q Q')$ be a state where $\CP$ is classical with alphabet $\alphCP$. A \term{tradeoff function on $\CP$} is a function $f: \alphCP \to \mathbb{R}$; equivalently, it is a real-valued tuple $\mbf{f} \in \mathbb{R}^{|\alphCP|}$ whose entries specify the values $f(\cP)$. For $\alpha \in (0,1) \cup (1,\infty)$, the \term{$f$-weighted entropy of order $\alpha$} of $\rho$ is
\begin{align}
\Halpha[f](Q|\CP Q')_{\rho}
&\coloneqq \frac{\alpha}{1-\alpha} \log \left( \sum_{\cP} \rho(\cP) \, 2^{\left(\frac{1-\alpha}{\alpha}\right) \left(\Halpha(Q|Q')_{\rho_{|\cP}} - f(\cP) \right)} \right),
\end{align}
where $\Halpha$ denotes the {\Renyi} conditional entropy \cite{Muller_Lennert_2013,WWY14}.
\end{definition}
 The inequality $\Halpha[f](Q|\CP Q')_{\rho} \geq 0$ can be interpreted as $f(\cP)$ being a lower bound on $\Halpha(Q|Q')_{\rho | \cP}$ in the log-mean-exponential sense (see~\cite[Sec~4.1]{arqand_marginal_2025}). Note that the $f$-weighted {\Renyi} entropy reduces to the usual {\Renyi} entropy when the function $f$ is set to be zero for all inputs \cite[Proposition 5.1]{tomamichel_quantum_2016}. In this work, we will need the following properties of the $f$-weighted  sandwiched {\Renyi} conditional entropies.

\begin{lemma}\label{lemma:DPIfweighted}
	(Data-processing \cite[Lemma~4.4]{arqand_marginal_2025}) Let $\rho \in \dop{=}(\CP Q Q')$ be classical on $\CP$, let $f$ be a tradeoff function on $\CP$, and take any $\alpha\in [\frac{1}{2},1)\cup(1,\infty]$. Then for any channel $\mathcal{E} \in \CPTP(Q',Q'')$, 
	\begin{align}
		\Halpha[f](Q|\CP Q'')_{\mathcal{E}[\rho]} \geq \Halpha[f](Q|\CP Q')_{\rho}.
	\end{align}
	If $\mathcal{E}$ is an isometry, then we have equality in the above bound.
\end{lemma}

\begin{lemma}(\cite[Lemma 6.4]{BDR3})\label{lemma:Monotonicity_f}
	(Monotonicity in $f$) Let $\rho \in \dop{=}(\CP Q Q')$ be classical on $\CP$, let $f,g$ be tradeoff functions on $\CP$ such that for all $\cP\in\CP$ we have $f(\cP)\leq g(\cP)$. Then, for any $\alpha\in [\frac{1}{2},1)\cup(1,\infty]$, we have
	\begin{align}
		\Halpha[g](Q|\CP Q')_{\rho} \leq \Halpha[f](Q|\CP Q')_{\rho}.
	\end{align}
\end{lemma}

We will also need the following definition in our proofs. 

\begin{definition}\label{def:purify}
For registers $Q,Q'$ with $\dim(Q) \leq \dim(Q')$, a purifying function for $Q$ onto $Q'$ is a function $\pf: \dop{\leq}(Q) \to \dop{\leq}(QQ')$ such that for any state $\rho_Q$, the state $\pf(\rho_Q)$ is a purification of $\rho_Q$ onto the register $Q'$, i.e.~a (possibly subnormalized) rank-$1$ operator such that $\tr_{Q'}[{\pf(\rho_Q)}] = \rho_Q$.
\end{definition}
Note that a purifying function is not a CPTP map.

\subsection{MEAT and QKD}

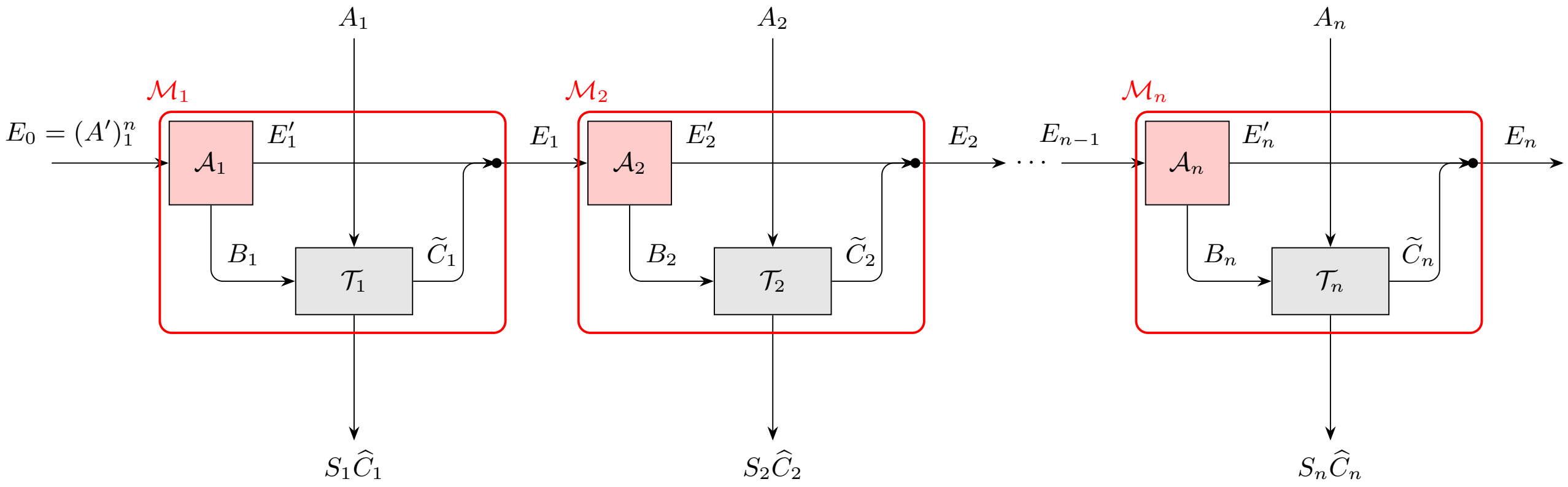
\begin{figure}[ht]
    \centering
    \scalebox{0.9}{%
\begin{tikzpicture}[>=Stealth,auto,
  box/.style  ={draw,minimum width=1cm,minimum height=1cm,
                align=center,fill=red!20},
  cpbox/.style ={draw,minimum width=1.4cm,minimum height=0.8cm,
                align=center,fill=gray!20},
  sig/.style  ={->,rounded corners=4pt},
  merge/.style={circle,fill,inner sep=1.2pt}
]

\node[box]   (N1) {$\attack{1}$};
\node[cpbox, below right=\offsetY and \offsetX of N1] (L1) {$\QKDGmap{1}$};

\node[box,   right=\colsep of N1] (N2) {$\attack{2}$};
\node[cpbox, below right=\offsetY and \offsetX of N2] (L2) {$\QKDGmap{2}$};

\node[right=\colsep of N2] (dots) {$\cdots$};
\node[box,   right= of dots] (Nn) {$\attack{n}$};
\node[cpbox, below right=\offsetY and \offsetX of Nn] (Ln) {$\QKDGmap{n}$};

\node[left =1.4cm of N1] (E0) {};
\node[right=4cm   of Nn] (En) {};

\coordinate (join1) at ($(L1.east |- N1) + (1.0,0)$);
\coordinate (join2) at ($(L2.east |- N2) + (1.0,0)$);
\coordinate (joinn) at ($(Ln.east |- Nn) + (1.0,0)$);

\draw[sig] (E0) node[above=2pt, xshift=10pt]{$E_{0}=(A')_1^n$} -- (N1.west);

\draw[sig] (N1.east) node[above=2pt, xshift=10pt] {$E'_1$} -- (join1);
\draw[sig] (L1.east) node[above=2pt, xshift=10pt]   {$\CPhat_{1}$}
           -- ++(0.6,0) |- (join1);
\node[merge] at (join1) {};
\draw[sig] (join1) -- node[above=2pt,xshift=1pt] {$E_{1}$} (N2.west);

\draw[sig] (N2.east) node[above=2pt, xshift=10pt] {$E'_2$} -- (join2);
\draw[sig] (L2.east) node[above=2pt, xshift=10pt]   {$\CPhat_{2}$}
           -- ++(0.6,0) |- (join2);
\node[merge] at (join2) {};
\draw[sig] (join2) -- node[above=2pt,xshift=1pt] {$E_{2}$} (dots.west);

\draw[sig] (Nn.east) node[above=2pt, xshift=10pt] {$E'_n$} -- (joinn);
\draw[sig] (Ln.east) node[above=2pt, xshift=10pt]   {$\CPhat_{n}$}
           -- ++(0.6,0) |- (joinn);
\node[merge] at (joinn) {};

\draw[sig] (dots.east) node[above=2pt,xshift = 3pt] {$E_{n-1}$} -- (Nn.west);
\draw[sig] (joinn) -- node[above=2pt,xshift=1pt] {$E_{n}$} (En);

\foreach \i/\N/\L in {1/N1/L1, 2/N2/L2, n/Nn/Ln}{
  \draw[sig] (\N.south) -- ++(0,-0.6)
             |- node[above right=3pt] {$B_{\i}$} (\L.west);
}

\foreach \i/\L in {1/L1, 2/L2, n/Ln}{
  \node[below=1.5cm of \L] (S\i) {$S_{\i}\widehat{C}_{\i}$};
  \draw[sig] (\L.south) -- (S\i);

  \node[above=2.5cm of \L] (A\i) {$A_{\i}$};
  \draw[sig,<-] (\L.north) -- (A\i);
}

\draw[red,thick,rounded corners]
  ($ (N1.north west) + (-3pt,3pt)$)
  rectangle
  ($ (join1 |- L1.south) + (3pt,-6pt)$);
\node[text=red] at ($($(N1.north west) + (0pt, 10pt)$)$) {\(\mathcal{M}_1\)};

\draw[red,thick,rounded corners]
  ($ (N2.north west) + (-3pt,3pt)$)
  rectangle
  ($ (join2 |- L2.south) + (3pt,-6pt)$);
\node[text=red] at ($($(N2.north west) + (0pt, 10pt)$)$) {\(\mathcal{M}_2\)};

\draw[red,thick,rounded corners]
  ($ (Nn.north west) + (-3pt,3pt)$)
  rectangle
  ($ (joinn |- Ln.south) + (3pt,-6pt)$);
\node[text=red] at ($($(Nn.north west) + (0pt, 10pt)$)$) {\(\mathcal{M}_n\)};
\end{tikzpicture}}
    \caption{Evolution of the state under Eve's attack channels $\attack{j}$ and the protocol maps $\QKDGmap{j}$ for the QKD protocol after source-replacement. Alice prepares the state $\sourcesymbol_{\Ameas_1^n \Ashield_1^n (A')_1^n}$ and sends $(A')_1^n$ to Eve, which becomes $E_0$. Each round consists of an attack $\attack{j} \in \CPTP(E_{j-1}, B_j E'_j)$ followed by the protocol map $\QKDGmap{j} \in \CPTP(A_j B_j, S_j \CP_j \CPhat_j)$, with $A_j = \Ameas_j \Ashield_j$. The classical announcement copy $\CPhat_j$ is merged with $E'_j$ to form Eve's updated side-information $E_j$. The combined per-round map is $\onerndchann_j = \QKDGmap{j} \circ \attack{j}$.}
    \label{fig:meat_attack}
\end{figure}

 As argued in \cite[Section 8.1]{BDR3}, the main goal during the security analysis of a QKD protocol is  analyzing the output of a sequence of CPTP maps, which represent Eve's attack and Alice's and Bob's operations in each round. These maps act on the so-called source-replaced state, which equivalently represents Alice's state preparation as measurements on an entangled state. The state evolution is depicted in \cref{fig:meat_attack}, and is understand as follows.

One typically considers a protocol where state preparation is independent in each round. Then, the global source replaced state is tensor product across the rounds, and is given by
\begin{equation} \label{eq:sourcerplacedstate_basic}
    \sourcesymbol^\mathrm{ind}_{\Ameas_1^n \Ashield_1^n (A')_1^n} = \bigotimes_{j=1}^n \sigma^{(j)}_{A_j \Ashield_j (A')_j},
\end{equation}
where we have the register $\Ameas_j$ (which Alice measures to obtain her classical setting choice), the shield register $\Ashield_j$ (which holds purifications of the prepared signal state), and the transmitted register $A'_j$, with $A_j \defvar \Ameas_j \Ashield_j$ denoting the registers which never leave Alice's lab.  The fact that this state is in tensor product form over the $A_1^n$ registers is  crucial to allow the application of MEAT to QKD, as will see below, and is violated when one considers correlated imperfections in source preparation, which we will see in \cref{sec:source-correlations}. Eve holds $E_j$ as her side-information registers after round $j$, and we set $E_0 = (A')_1^n$. 

In each round $j$, Eve first implements her attack for that round, which is given by an arbitrary channel $\attack{j} \in \CPTP(E_{j-1}, B_j E'_j)$, taking her current side information $E_{j-1}$ and producing both the register $B_j$ that she forwards to Bob and an updated side-information register $E'_j$ that she keeps. This is followed by Alice's and Bob's operations for that round, described by the ``protocol map" $\QKDGmap{j} \in \CPTP(A_j B_j, S_j \CP_j \CPhat_j)$. This captures the following sub-steps in order: (i) Alice's measurement on $\Ameas_j$ (in the source-replaced picture) and Bob's measurement on $B_j$, which produce classical outcomes stored in $X_j$ and $Y_j$, respectively; (ii) the public-announcement step, which generates $\CP_j$ as a classical function of $X_jY_j$ and releases a classical copy $\CPhat_j$ to Eve; and (iii) the sifting/key-map step, which produces the per-round secret value $S_j$ from $X_j$ and $\CP_j$ on Alice's side. The registers $X_jY_j$ are discarded at the output of $\QKDGmap{j}$ because they are not needed for the subsequent analysis. The global state at the end of this sequence of maps is given by
\begin{equation}
    \rho_{S_1^n \CP_1^n E_n} = \left(  \QKDGmap{n} \circ \attack{n} \dots \QKDGmap{1} \circ \attack{1} \right)\!\left[\sourcesymbol^\mathrm{ind}_{A_1^n (A')_1^n}\right].
\end{equation}
Note that we will typically assume that there is some notion of ``test" and ``generation" rounds. In the test rounds, Alice and Bob typically announce the exact state sent and measurement outcome obtained, and store the value $\bot$ in $S_j$. In key generation rounds, Alice and Bob only announce partial information (such as basis choice), and store bits in the $S_j$ register. 

Let us consider a fixed attack $\{ \attack{j} \}$ by the adversary. Our goal is to construct a global tradeoff function $\fhatfull$ on $\CP_1^n$ such that
\begin{equation}\label{eq:globaltradeoff-goal}
    \Halpha[\fhatfull](S_1^n | \CP_1^n E_n)_\rho \geq 0
\end{equation}
holds for the state $\rho_{S_1^n \CP_1^n E_n}$ produced at the end of the sequence of channels described above. The bound in \cref{eq:globaltradeoff-goal} is equivalent to having $\fhatfull(\cP_1^n)$ lower bound $\Halpha(S_1^n | E_n)_{\rho_{|\cP_1^n}}$ in the log-mean-exponential sense (\cref{def:fweighted}). The following theorem allows us to obtain such a tradeoff function.

\begin{theorem}(MEAT \cite[Theorem 4.1a]{arqand_marginal_2025} applied to QKD \cite[Theorem 8.1]{BDR3})
\label{thm:meatqkd}
Consider the sequence of channels in \cref{fig:meat_attack}, where we have
$\rho_{S_1^n \CP_1^n E_n} = \left(  \QKDGmap{n} \circ \attack{n} \dots \QKDGmap{1} \circ \attack{1} \right)\!\left[\sourcesymbol^\mathrm{ind}_{A_1^n (A')_1^n}\right]$, and $\sourcesymbol^\mathrm{ind}_{A_1^n} = \bigotimes_{j=1}^n \sigma^{(j)}_{A_j}$ is tensor product on the $A_1^n$ systems (\cref{eq:sourcerplacedstate_basic}). For each $j$ and each value of $\cP_1^{j-1}$, let $f_{|\cP_1^{j-1}}$ be a tradeoff function on $\CP_j$. Define the single-round set 
\begin{equation}
\Sigma_j(\attack{j}) \defvar \left\{
\QKDGmap{j} \circ \attack{j} [\omega_{A_j E_{j-1} \widetilde{E}}]
\;\middle|\;
\omega \in \dop{=}(A_j E_{j-1} \widetilde{E}), \;\omega_{A_j} = \sigmaconstraint{j}
\right\},
\end{equation}
where $\widetilde{E}$ is a purifying register large enough to purify $A_j E_{j-1}$, and the per-round quantity
\begin{equation} \label{eq:kappabasic}
\kappafuncgeneric{f_{|\cP_1^{j-1}}}{\sigmaconstraint{j}}{\QKDGmap{j}}{\attack{j}}\defvar
\inf_{\nu \in \Sigma_j(\attack{j})} \Halpha[f_{|\cP_1^{j-1}}](S_j | \CP_j E_j \widetilde{E})_\nu .
\end{equation}
Then for any $\alpha \in (1,\infty)$, the function
\begin{equation}
\fhatfull(\cP_1^n) \defvar \sum_{j=1}^n \left(f_{|\cP_1^{j-1}}(\cP_j) + \kappafuncgeneric{f_{|\cP_1^{j-1}}}{\sigmaconstraint{j}}{\QKDGmap{j}}{\attack{j}}\right)
\end{equation}
satisfies $\Halpha[\fhatfull](S_1^n | \CP_1^n E_n)_\rho \geq 0$.
\end{theorem}

Thus, the application of MEAT directly gives us an appropriate global tradeoff function $\fhatfull$, in terms of the single-round quantities $\kappafuncgeneric{f_{|\cP_1^{j-1}}}{\sigmaconstraint{j}}{\QKDGmap{j}}{\attack{j}}$. The final remaining step is to additionally minimize the $\kappa$ term over all such attacks channels, in order to obtain a tradeoff function that is valid for arbitrary attacks. This poses no additional difficulty, and numerical methods to do so have already been developed in \cite{kamin_renyi_2025,navarro_finite_2025}. In the remainder of this work, we focus on obtaining the required $\fhatfull$ (in terms of suitable single-round quantities). We start with detector correlations and iterative sifting.

\section{MEAT for detector correlations}
\label{sec:detector-memory}
The protocol map $\QKDGmap{j}$ in \cref{fig:meat_attack} acts only on $A_j$ (Alice's source-replaced register for round $j$) and $B_j$ (the register Eve forwards to Bob in round $j$); it has no input from any register that persists across rounds. As a consequence, the model of \cref{fig:meat_attack} cannot capture device behaviours in which a detection event in one round influences subsequent rounds through the internal state of the detector. Thus, we cannot model well-known examples of correlated detector imperfections, such as afterpulsing or dead times.

\begin{figure}[ht]
    \centering
    \scalebox{0.9}{%
\begin{tikzpicture}[>=Stealth,auto,
  box/.style  ={draw,minimum width=1cm,minimum height=1cm,
                align=center,fill=red!20},
  cpbox/.style ={draw,minimum width=1.4cm,minimum height=0.8cm,
                align=center,fill=gray!20},
  sig/.style  ={->,rounded corners=4pt},
  merge/.style={circle,fill,inner sep=1.2pt}
]

\node[box]   (N1) {$\attack{1}$};
\node[cpbox, below right=\offsetY and \offsetX of N1] (L1) {$\QKDGmap{1}$};

\node[box,   right=\colsep of N1] (N2) {$\attack{2}$};
\node[cpbox, below right=\offsetY and \offsetX of N2] (L2) {$\QKDGmap{2}$};

\node[right=\colsep of N2] (dots) {$\cdots$};
\node[box,   right= of dots] (Nn) {$\attack{n}$};
\node[cpbox, below right=\offsetY and \offsetX of Nn] (Ln) {$\QKDGmap{n}$};

\node[left =1.4cm of N1] (E0) {};
\node[right=4cm   of Nn] (En) {};

\coordinate (join1) at ($(L1.east |- N1) + (1.0,0)$);
\coordinate (join2) at ($(L2.east |- N2) + (1.0,0)$);
\coordinate (joinn) at ($(Ln.east |- Nn) + (1.0,0)$);

\draw[sig] (E0) node[above=2pt, xshift=10pt]{$E_{0}=(A')_1^n$} -- (N1.west);

\draw[sig] (N1.east) node[above=2pt, xshift=10pt] {$E'_1$} -- (join1);
\draw[sig] (L1.east) node[above=2pt, xshift=10pt]   {$\CPhat_{1}$}
           -- ++(0.6,0) |- (join1);
\node[merge] at (join1) {};
\draw[sig] (join1) -- node[above=2pt,xshift=1pt] {$E_{1}$} (N2.west);

\draw[sig] (N2.east) node[above=2pt, xshift=10pt] {$E'_2$} -- (join2);
\draw[sig] (L2.east) node[above=2pt, xshift=10pt]   {$\CPhat_{2}$}
           -- ++(0.6,0) |- (join2);
\node[merge] at (join2) {};
\draw[sig] (join2) -- node[above=2pt,xshift=1pt] {$E_{2}$} (dots.west);

\draw[sig] (Nn.east) node[above=2pt, xshift=10pt] {$E'_n$} -- (joinn);
\draw[sig] (Ln.east) node[above=2pt, xshift=10pt]   {$\CPhat_{n}$}
           -- ++(0.6,0) |- (joinn);
\node[merge] at (joinn) {};

\draw[sig] (dots.east) node[above=2pt,xshift = 3pt] {$E_{n-1}$} -- (Nn.west);
\draw[sig] (joinn) -- node[above=2pt,xshift=1pt] {$E_{n}$} (En);

\foreach \i/\N/\L in {1/N1/L1, 2/N2/L2, n/Nn/Ln}{
  \draw[sig] (\N.south) -- ++(0,-0.6)
             |- node[above right=3pt] {$B_{\i}$} (\L.west);
}

\foreach \i/\L in {1/L1, 2/L2, n/Ln}{
  \node[below=1.5cm of \L] (S\i) {$S_{\i}\widehat{C}_{\i}$};
  \draw[sig] (\L.south) -- (S\i);

  \node[above=2.5cm of \L] (A\i) {$A_{\i}$};
  \draw[sig,<-] (\L.north) -- (A\i);
}

\draw[red,thick,rounded corners]
  ($ (N1.north west) + (-3pt,3pt)$)
  rectangle
  ($ (join1 |- L1.south) + (3pt,-6pt)$);
\node[text=red] at ($($(N1.north west) + (0pt, 10pt)$)$) {\(\mathcal{M}_1\)};

\draw[red,thick,rounded corners]
  ($ (N2.north west) + (-3pt,3pt)$)
  rectangle
  ($ (join2 |- L2.south) + (3pt,-6pt)$);
\node[text=red] at ($($(N2.north west) + (0pt, 10pt)$)$) {\(\mathcal{M}_2\)};

\draw[red,thick,rounded corners]
  ($ (Nn.north west) + (-3pt,3pt)$)
  rectangle
  ($ (joinn |- Ln.south) + (3pt,-6pt)$);
\node[text=red] at ($($(Nn.north west) + (0pt, 10pt)$)$) {\(\mathcal{M}_n\)};

\draw[sig] ([xshift=-1.2cm, yshift=-8pt]L1.west)
           node[above=2pt, xshift=10pt]{$\Mem_{0}$}
           -- ([yshift=-8pt]L1.west);

\draw[sig] ([yshift=-8pt]L1.east) -- node[above=2pt]{$\Mem_{1}$}
           ([yshift=-8pt]L2.west);

\node (mdots) at ($([yshift=-8pt]L2.east)!0.5!([yshift=-8pt]Ln.west)$) {$\cdots$};
\draw[sig] ([yshift=-8pt]L2.east) -- node[above=2pt]{$\Mem_{2}$} (mdots.west);
\draw[sig] (mdots.east) -- node[above=2pt]{$\Mem_{n-1}$} ([yshift=-8pt]Ln.west);

\draw[sig] ([yshift=-8pt]Ln.east) -- ++(0.9,0)
           node[above=2pt, xshift=-3pt]{$\Mem_{n}$};
\end{tikzpicture}}
    \caption{The sequence of channels of \cref{fig:meat_attack} augmented with a device-memory register $\Mem_j$ that flows from $\QKDGmap{j}$ into $\QKDGmap{j+1}$. Each protocol map is now $\QKDGmap{j} \in \CPTP(A_j B_j \Mem_{j-1}, S_j \CP_j \CPhat_j \Mem_j)$. The initial register $\Mem_0$ is a fixed state (e.g.\ a fresh-detector state) and the final register $\Mem_n$ is discarded.}
    \label{fig:meat_attack_memory}
\end{figure}
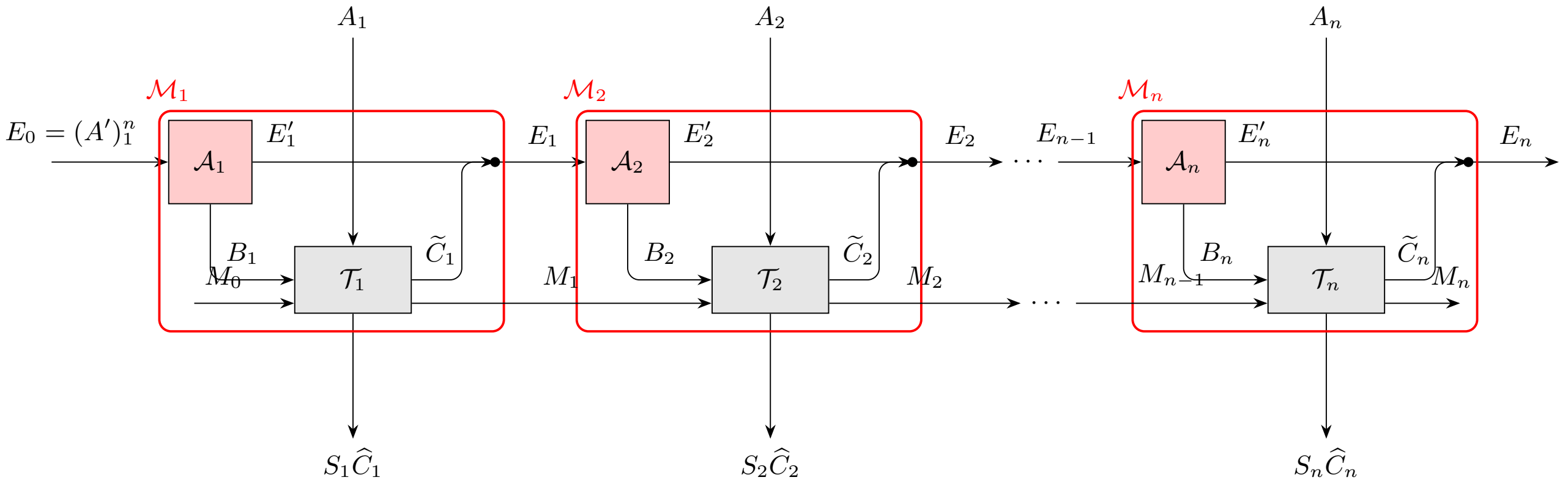

To accommodate such effects within our model, we include an additional memory register $\Mem_j$ that flows from one round to the next, as depicted in \cref{fig:meat_attack_memory}.  The register $\Mem_j$ represents \emph{device memory}, that is, internal state of Alice's and Bob's apparatus\footnote{In the case of entanglement-based protocols, both Alice and Bob use detectors.} that persists across rounds, and can model afterpulsing, dead times, and similar correlated detector effects. (Iterative sifting can also be modeled using this approach, as we shall see later). Concretely, the protocol map in round $j$ now has an additional input register $M_{j-1}$ and produced an additional output register $M_j$, and $\QKDGmap{j} \in \CPTP(A_j B_j \Mem_{j-1}, S_j \CP_j \CPhat_j \Mem_j)$, with $\Mem_0$ a fixed initial state.

In our analysis of the scenario from \cref{fig:meat_attack_memory}, the following aspects are important to note. 

\begin{enumerate}
    \item The MEAT statement does not allow for a private memory register to model internal device state; registers that pass between rounds have to be included in Eve's side-information. Therefore, in our analysis, we will be forced to give Eve access to $\Mem_j$. Thus, we will not be able to meaningfully analyze scenarios where, in order to implement the correct correlated behaviour in round $j$, we have to include excessive information in $\Mem_{j-1}$. 
    
    As a concrete example, consider dead times: implementing the correct POVM in round $j$ depends on the exact outcome of Bob's measurement in the previous round (since the appropriate detector needs to be shut off). If this information is included in $\Mem_{j-1}$, then it is also given to Eve, who will know Bob's measurement outcome, and therefore his final key. 

   We will remedy this as follows. We will assume correlated effects persist for some finite number of rounds $l^\mathrm{det}_c$ after a detection event. In these rounds, we will assume that all detectors are switched ``off" (via hardware modifications) and cannot register a click, and Alice and Bob simply record a special symbol $\bot^\mathrm{discard}$ in the public announcement register $\CP_j$ and store $\bot$ in the key register $S_j$. This effectively ensures that we do not need to know the exact click pattern of previous rounds to describe the operations of round $j$: we only need to know whether previous rounds had detections or no-detections, which is information that is typically announced anyway in most QKD protocols.

    \item The MEAT statement can be directly applied to the sequence of channels in \cref{fig:meat_attack_memory}, with $E_{j-1}$ replaced by $E_{j-1}\Mem_{j-1}$. The apparent difficulty is that the corresponding single-round optimization contains the possibly large register $\Mem_{j-1}$. Note that even if we only store detect vs no-detect information in the memory, $\Mem_{j-1}$ would need to store information pertaining to the closest detection event in history, and thus has dimension at least $l^\mathrm{det}_c$.
    
    However, the behaviour of the protocol map only depends on much more coarse-grained information from $\Mem_{j-1}$. In the example described above, the protocol map $\QKDGmap{j}$ does trivial operations when $\Mem_{j-1}$ indicates that there was a detection in any of the previous $l^\mathrm{dec}_c$ rounds, and implements the usual measurements and public announcements if there were no detections. Thus, it is reasonable to attempt to replace the $\Mem_{j-1},\Mem_j$ registers which appear in the naive formulation of the single-round optimization $\kappa$ (\cref{eq:kappabasic}), with smaller registers that only record the relevant coarse-grained memory. This is important to render the entire calculation numerically feasible. 
    \item The single-round quantities we get from MEAT involve an optimization over all the input states $\omega_{A_j M_{j-1} E_{j-1} \widetilde{E}}$ satisfying $\omega_{A_j}=\sigma^{(j)}_{A_j}$. Since we have no constraint on $M_{j-1}$, and the protocol map deterministically discards rounds for certain values of $M_{j-1}$, a naive application would result in the infimum being obtained for the $\omega_{A_j M_{j-1} E_{j-1} \widetilde{E}}$ that deterministically leads to trivial operations.
    
    To remedy this, we design the protocol maps so that the relevant memory effect is indicated through the public announcements, and therefore also included in the test announcements. This rules out the infimum being obtained for $\omega_{A_j M_{j-1} E_{j-1} \widetilde{E}}$.
    \footnote{This is a lot easier to see when considering the fixed-length variation of the MEAT result, which explicitly includes an optimization over states where the marginal on the $\CP$ belongs to the predetermined set of accepted statistics. In the variable length case, this is manifested by the fact that the tradeoff function can now assign different scores to rounds that were discarded normally, and rounds that were discarded due to different memory effects.  } See \cite[Section 5]{nahar2025imperfect}, \cite[Chapter 7]{nahar_phd_2025}, where many of these ideas were first sketched out.  
\end{enumerate}

Having explained the motivation behind the approach taken in this work, we now start working towards obtaining a variant of the MEAT that is applicable to QKD with detector correlations. Note that while it is useful to keep the above considerations in mind, the remainder of this section can be read on its own, without referring to any specific protocol modification or model for detector correlations.

\begin{remark} \label{remark:conditioning_imperfections}
Note that the GEAT \cite{metger_generalised_2022} and GREAT \cite{arqand_generalized_2024} both allow for a secret memory register. Moreover, the secret memory register needs to satisfy the ``no-signaling" condition, i.e., $M_{j-1}$ cannot signal to $E_j$. Since we wish to use the memory register to model correlated detector imperfections, $M_{j-1}$ affects the POVMs used during measurements and therefore affects the announcements performed (which are included in $E_j$). Thus, while it may be tempting to utilize the secret memory register within those theorems, or hope to include such a register within MEAT (along with a no-signaling condition), the no-signaling condition typically rules out a straightforward resolution to the problem of correlated detector imperfections. 
\end{remark}

\subsection{The setting}
We will now describe the conditions that must hold for the protocol maps $\QKDGmap{j}$, for our results to be applicable. These conditions are motivated by the discussion above.

\begin{condition}[Memory with coarse-gaining] \label{condition:coarsegrainingmemory} We let $M_{j-1}$ be the classical memory register that is input to round $j$. Its alphabet may be large and may, for instance, contain the complete detection pattern for all previous rounds. However, the protocol map only depends on some coarse-grained information contained in $M_{j-1}$. This can be represented as some system $\EffMem_{j}$ that is created via a map 
    $\coarsegrainedhistory{j} \in \CPTP(M_{j-1}, M_{j-1} \EffMem_{j})$. This map simply looks at $M_{j-1}$ and stores the coarse-grained value of the memory in $\EffMem_j$, and does not ``disturb" the $M_{j-1}$ system. I.e, we have
\begin{equation}
    \coarsegrainedhistory{j} \left[ \ketbra{m_{j-1}}_{M_{j-1}} \right] =  \ketbra{m_{j-1}}_{M_{j-1}}\ketbra{f_\mathrm{coarse} (m_{j-1})}_{\EffMem_{j}} \qquad \qquad \forall m_{j-1} 
\end{equation}
\end{condition}

\begin{condition}[Classical Memory and protocol maps] \label{condition:classicalmemoryandmap} We then consider a CPTP map $\QKDGmap{j}^{(\effmem_j)} \CPTP(A_j B_j, S_j \CP_j \CPhat_j)$ that is applied whenever the $\EffMem_j$ register stores $d_j$. This can be represented further by the map $   \effectiveQKDGmap{j} \in \CPTP(A_j B_j \EffMem_j, S_j \CP_j \CPhat_j)$:

\begin{equation}
\effectiveQKDGmap{j} = \Tr_{\EffMem_j} \circ \left( \sum_{\effmem_j} \ketbra{\effmem_j}_{\EffMem_j} \otimes \QKDGmap{j}^{(\effmem_j)}   \right)
\end{equation}
\end{condition}

\begin{condition}[Creating the next memory register] \label{condition:creatingnextmemory}This is then followed by a map $\createhistory{j} \in \CPTP(\CP_j M_{j-1} , \CP_j M_j)$ that updates the memory register $M_j$ based on the previous history and public announcements of that round. We require that $\createhistory{j}$ does not ``disturb" the $\CP_j$ register: It simply reads the value $\cP_j$, and then applies a function $f^\mathrm{update}_{\cP_j} (m_{j-1}) = m_j$ to produce the value stored in $m_j$. Thus, we have
\begin{equation}
    \createhistory{j} \left[ \ketbra{\cP_j}_{\CP_j} \otimes \ketbra{m_{j-1}}_{M_{j-1}}\right] = \ketbra{\cP_j}_{\CP_j} \otimes \ketbra{ f^\mathrm{update}_{\cP_j} (m_{j-1})}_{M_j}
\end{equation}

\end{condition}

Thus, in each round, Alice and Bob implement a channel $\QKDGmap{j} \in \CPTP(A_j B_j M_{j-1}, S_j \CP_j \CPhat_j M_j)  $ that is given by
\begin{equation}
    \QKDGmap{j} = \createhistory{j} \circ \effectiveQKDGmap{j} \circ \coarsegrainedhistory{j}
\end{equation}
    
\begin{condition}[Public announcements indicate memory effect] \label{condition:publicannouncementsindicatememory} We now impose the constraint that the public announcements are different for different values in $\EffMem_j$. That is, let $\{\projmap_{\CP_j}^{(\effmem_j)}\}_{\effmem_j\in\effmemSet_{j}}$ be pairwise orthogonal projectors that are diagonal in the classical basis of $\CP_j$. Then, we have that
    \begin{equation}
\begin{aligned}
\projmap_{\CP_j}^{(\effmem_j)}  \QKDGmap{j}^{(\effmem_j)} 
[ \omega]  \projmap_{\CP_j}^{(\effmem_j)} = \QKDGmap{j}^{(\effmem_j)} [ \omega], \qquad 
\qquad\forall \omega \in \dop{A_j B_{j} M_{j-1} } .
\end{aligned}
\end{equation}
\end{condition}

\subsection{Proving the MEAT variant}
As argued earlier, we are free to apply the MEAT statement to the scenario from \cref{fig:meat_attack_memory}, and treat the $\Mem_j$ as being accessible to Eve. We thus focus on the resulting single-round $\kappa$ quantity that is obtained from the application of the MEAT. We will first prove \cref{lem:effective-history-reduction}, which reduces the size of the computation we would have to perform to compute this single-round quantity. In particular, will be able to replace the $M_{j-1}$ register in the optimization with the much smaller $\EffMem_j$.
Furthermore, we will be able to entirely remove the $M_j$ system on the output side, since it can be created from $\CP_j$ (which is known to Eve), and $M_{j-1}$ (a purification of which is present in the purifying system $\widetilde{E}$ which appears in the single round $\kappa$).

\begin{lemma}[Reducing to coarse-grained memory registers]
\label{lem:effective-history-reduction}
Consider the scenario in \cref{fig:meat_attack_memory}, and suppose \cref{condition:classicalmemoryandmap,condition:creatingnextmemory,condition:coarsegrainingmemory} are satisfied, the source-replaced state is given by \cref{eq:sourcerplacedstate_basic}, and let $\alpha \in (1,\infty)$. Consider the single-round optimization that would be obtained through a direct application of the MEAT, for any tradeoff function $f$ on $\CP_j$, which is given by
\begin{equation} \label{eq:historykappadefined}
\begin{aligned}
     \kappafuncgeneric{f}{\sigmaconstraint{j}}{\QKDGmap{j}}{\attack{j}} &= \inf_{\nu\in\Sigma_j^{\mathrm{hist}}(\attack{j})}
\Halpha[f](S_j|\CP_j E_j\Mem_j\widetilde E)_\nu,  \\
\Sigma_j^{\mathrm{hist}}(\attack{j})
&\defvar\bigl\{\,
\QKDGmap{j}\circ
(\attack{j}\otimes\id_{\Mem_{j-1}})
[\omega_{A_jE_{j-1}\Mem_{j-1}\widetilde E}]
\;\big|\;
\omega\in\dop{=}(A_jE_{j-1}\Mem_{j-1}\widetilde E), \omega_{A_j} = \sourcesymbol^{(j)}_{A_j} \big\}
\end{aligned}
\end{equation}
where $\widetilde{E}$ is a purifying system.
Consider a related single-round optimization, given by
\begin{equation} \label{eq:effectivekappadefined}
\begin{aligned}
     \kappafuncgeneric{f}{\sigmaconstraint{j}}{\effectiveQKDGmap{j}}{\attack{j}} &= \inf_{\nu\in\Sigma_j^{\mathrm{eff}}(\attack{j})}
\Halpha[f](S_j|\CP_jE_j\widehat{E})_\nu,  \\
\Sigma_j^{\mathrm{eff}}(\attack{j})
&\defvar\bigl\{\,
\effectiveQKDGmap{j}\circ
(\attack{j}\otimes\id_{\EffMem_{j}})
[\omega_{A_jE_{j-1}\EffMem_{j}\widehat{E}}]
\;\big|\;
\omega\in\dop{=}(A_jE_{j-1}\EffMem_{j}\widehat{E}), \omega_{A_j} = \sourcesymbol^{(j)}_{A_j} \big\}
\end{aligned}
\end{equation}
where $\widehat{E}$ is a purifying system.
Then, 
\begin{equation}
     \kappafuncgeneric{f}{\sigmaconstraint{j}}{\effectiveQKDGmap{j}}{\attack{j}} \leq  \kappafuncgeneric{f}{\sigmaconstraint{j}}{\QKDGmap{j}}{\attack{j}}
\end{equation}

\end{lemma}
\begin{proof}
Let $\omega^*_{A_j E_{j-1} M_{j-1} \widetilde{E}}$ be any state in the feasible set of \cref{eq:historykappadefined}. It is enough to show that there exists a corresponding $\omega'_{A_j E_{j-1} \EffMem_{j} \widetilde{E}}$ in the set $\Sigma^\mathrm{eff}_j(\attack{j})$ from \cref{eq:effectivekappadefined}, which results in a lower value of the objective function in \cref{eq:effectivekappadefined}. We show that this is indeed the case. Without loss of generality we can restrict our attention to states classical in $\Mem_{j-1}$, and therefore the purifying system $\widetilde{E}$ stores a copy of $\Mem_{j-1}$. (Since the protocol map treats $M_{j-1}$ classically and Eve’s attack acts trivially on it, pinching this register leaves the output state, and hence the objective function, unchanged while preserving the marginal constraint on $A_j$. Thus the state obtained by giving Eve a purification after pinching the $
\Mem_{j-1}$ register, yields a lower value of the objective function via data processing (\cref{lemma:DPIfweighted}).)

Writing $\omega^*_{A_j B_j \Mem_{j-1} E_j' \widetilde{E}} = \attack{j} [\omega^*_{A_j E_{j-1} \Mem_{j-1} \widetilde{E}}] $, let 
\begin{equation}
\begin{aligned}
    \omega^*_{A_j E_{j-1} \EffMem_{j} \widetilde{E}} &=   \Tr_{\Mem_{j-1}} \circ \coarsegrainedhistory{j} [ \omega^*_{A_j E_{j-1} M_{j-1} \widetilde{E} } ]  \\
    \omega'_{A_j E_{j-1} \EffMem_{j} \widetilde{E} \widetilde{E}'} & \defvar  \pf \left( \Tr_{\Mem_{j-1}} \circ \coarsegrainedhistory{j} [ \omega^*_{A_j E_{j-1} M_{j-1} \widetilde{E}} ] \right) 
\end{aligned}
\end{equation}
where $\pf$ is a purifying function (\cref{def:purify}) that creates the $\widetilde{E}'$ register. Since $\coarsegrainedhistory{j} $ acts trivially on the $A_j$ system,   $\omega'_{A_j E_{j-1}
\EffMem_{j} \widetilde{E}\widetilde{E}'}$ satisfies the marginal constraint ($\omega'_{A_j} = \sigma^{(j)}_{A_j}$), and therefore belongs to $\Sigma^\mathrm{eff}_j(\attack{j})$. Thus we only need to show that  $\omega'_{A_j E_{j-1}
\EffMem_{j} \widetilde{E}\widetilde{E}'}$ leads to a lower value of the objective function in \cref{eq:effectivekappadefined}.

To show this, we have
\begin{align}
\Halpha[f](S_j|\CP_j E_j\Mem_j\widetilde E)_{  \createhistory{j} \circ \effectiveQKDGmap{j} \circ \coarsegrainedhistory{j} [ \omega^*_{A_j B_j \Mem_{j-1} \widetilde{E}}]  }   &=\Halpha[f](S_j|\CP_j E_j M_{j-1} \widetilde E)_{  \effectiveQKDGmap{j} \circ \coarsegrainedhistory{j} [ \omega^*_{A_j B_j \Mem_{j-1} \widetilde{E}}]  } \label{eq:kappalemmatemp2} \\
     &= \Halpha[f](S_j|\CP_j E_j  \widetilde E)_{\Tr_{M_{j-1}} \circ  \effectiveQKDGmap{j} \circ \coarsegrainedhistory{j} [ \omega^*_{A_j B_j \Mem_{j-1} \widetilde{E}}]  } \label{eq:kappalemmatemp3} \\
     &= \Halpha[f](S_j|\CP_j E_j  \widetilde E)_{  \effectiveQKDGmap{j} \circ \Tr_{M_{j-1}} \circ  \coarsegrainedhistory{j} [ \omega^*_{A_j B_j \Mem_{j-1} \widetilde{E}}].\label{eq:kappalemmatemp4}} \\
     &=\Halpha[f](S_j|\CP_j E_j  \widetilde E)_{  \effectiveQKDGmap{j} \circ \attack{j} [\omega^*_{A_j E_{j-1} \EffMem_{j-1} \widetilde{E}}].\label{eq:kappalemmatemp5}} \\
     &\geq  \Halpha[f](S_j|\CP_j E_j  \widehat E)_{  \effectiveQKDGmap{j} \circ \attack{j} [\omega'_{A_j E_{j-1} \EffMem_{j-1} \widehat{E} }].\label{eq:kappalemmatemp6}}
     \end{align}
  We use data processing (\cref{lemma:DPIfweighted}) in either direction to obtain \cref{eq:kappalemmatemp2}. One direction is trivial, and is obtained via $\createhistory{j}$. For the other direction,
  we use that fact that $\widetilde{E}$ stores a copy of the $\Mem_{j-1}$ register, and thus we can simply trace out $M_j$, and create $M_{j-1}$ by copying the relevant subregisters from the $\widetilde{E}$ system. We again use data processing in either direction to obtain \cref{eq:kappalemmatemp3}:  one direction is trivial (obtained via $\Tr_{\Mem_{j-1}}$), for the other direction, we can again create $M_{j-1}$ by copying the relevant subregisters from the $\widetilde{E}$ system. Finally, \cref{eq:kappalemmatemp4} follows from the fact that $\Tr_{\Mem_{j-1}}$ and $\effectiveQKDGmap{j}$  commute. We get \cref{eq:kappalemmatemp5} from the fact that Eve's attack $\attack{j}$ commutes with $ \Tr_{\Mem_{j-1}} \circ \coarsegrainedhistory{j}$. Finally, \cref{eq:kappalemmatemp6} follows from data processing, where $\widehat{E} = \widetilde{E} \widetilde{E}'$. In particular, it uses the fact that there exists a map from $\widehat{E}$ to $\widetilde{E}$ that maps $ \omega'_{A_j E_{j-1} \EffMem_{j} \widehat{E}}$ to $ \omega^*_{A_j E_{j-1} \EffMem_{j} \widetilde{E}}$, since the former state is a purification of the latter. 
\end{proof}

With this lemma, we can restrict our attention to computing the single-round quantity defined in \cref{eq:effectivekappadefined}. We can now state the result obtain by applying MEAT on \cref{fig:meat_attack_memory}.

\begin{theorem}[MEAT for detector correlations]
\label{thm:meat-effective-history} Consider the sequence of channels  in \cref{fig:meat_attack_memory}, where the output state is given by  $\rho_{S_1^n \CP_1^n M_n E_n} = \QKDGmap{n} \circ \attack{n} \dots  \QKDGmap{1} \circ \attack{1}  [\sourcesymbol^\mathrm{ind}_{A_1^n E_0}]$, and the input state satisfies $\sourcesymbol^\mathrm{ind}_{A_1^n} = \sigma^{(1)}_{A_1} \otimes \dots \otimes \sigma^{(n)}_{A_n}$. Suppose that the protocol maps $\QKDGmap{j}$ satisfy \cref{condition:classicalmemoryandmap,condition:creatingnextmemory,condition:coarsegrainingmemory} for all $j \in \{1,\dots,n\}$. For every value $j$, and every value of $\cP_1^{j-1}$, let $f_{|\cP_1^{j-1}}$ be a tradeoff function on $\CP_j$. For any $\alpha\in(1,\infty)$ define the global tradeoff 
function 
\begin{equation}
\label{eq:effective-history-global-tradeoff}
\begin{aligned}
\fhatfull(\cP_1^n)
\defvar
\sum_{j=1}^n\Bigl(&f_{|\cP_1^{j-1}}(\cP_j)
{}+\kappafuncgeneric{f_{|\cP_1^{j-1}}}{\sigmaconstraint{j}}{\effectiveQKDGmap{j}}{\attack{j}}\Bigr)
\end{aligned}
\end{equation}
where $\kappafuncgeneric{f_{|\cP_1^{j-1}}}{\sigmaconstraint{j}}{\effectiveQKDGmap{j}}{\attack{j}}$ is as defined in \cref{eq:effectivekappadefined}. Then, 
\begin{equation}
\label{eq:effective-history-meat-conclusion}
    \Halpha[\fhatfull](S_1^n|\CP_1^nE_n)_\rho
    \geq
    \Halpha[\fhatfull](S_1^n|\CP_1^nE_n\Mem_n)_\rho
    \geq0.
\end{equation}
\end{theorem}
\begin{proof}
The proof follows in two simple steps. First we apply \cref{thm:meatqkd}, which is simply a direct application of the MEAT \cite{arqand_marginal_2025} for the setting considered in \cref{fig:meat_attack_memory}, with $\Mem_{j-1} E_{j-1}$ as Eve's side-information register.  This directly yields a $\fhatfull$ satisfying \cref{eq:effective-history-meat-conclusion}, where the single-round optimization is given by \cref{eq:historykappadefined}. The second step is the application of \cref{lem:effective-history-reduction} and replacing the single-round optimization with the \cref{eq:effectivekappadefined}.
\end{proof}

Note that we have stated the above results without imposing \cref{condition:publicannouncementsindicatememory}. This is because the statements themselves are valid without that condition: however, \cref{condition:publicannouncementsindicatememory} will be necessary to get non-trivial results in all the concrete scenarios we will consider in \cref{subsec:detectorexamples}, and we defer a detailed discussion of this aspect to that subsection.

\subsubsection{Further improvements for entanglement-based protocols}

At this point, one may be tempted to look at the single-round optimization from \cref{eq:effectivekappadefined}, and attempt to simplify it further,  such that each memory effect (determined by the classical value of $\EffMem_j$) is analyzed ``separately" in some sense. We will now show that this can indeed be done (with \cref{condition:publicannouncementsindicatememory}), but it results in the loss of the marginal constraint in the resulting single-round $\kappa$ optimization. Intuitively, this is because while we can impose the marginal constraint on $\omega_{A_j \EffMem_{j} E_{j-1} \widetilde{E}} = \sum_{\effmem_j}  \Pr(\Omega(\effmem_j)) \ketbra{\effmem_j}_{\EffMem_j} \otimes \omega_{A_j  E_{j-1} \widetilde{E} | \Omega(\effmem_j)}$, the states $\omega_{A_j  E_{j-1} \widetilde{E} | \Omega(\effmem_j)}$ (where $\Omega(\effmem_j)$ denotes the event that $\EffMem_j$ stores $\effmem_j$), do not necessarily satisfy the same constraint. The loss of marginal constraint limits the applicability of the following results to entanglement-based protocols, and prohibits their application to prepare-and-measure protocols. Note that since we cannot retain the marginal constraint on $A_j$, we choose to omit that register when stating the results below. In the following statement, it is helpful to think of the $B_j$ register as storing the quantum systems that Eve sends to both Alice and Bob, in an entanglement-based protocol.

\begin{lemma}[Reducing to separate optimizations per memory]
\label{lem:effective-history-reduction_nomarginal}
Consider the scenario in \cref{fig:meat_attack_memory}, and suppose \cref{condition:classicalmemoryandmap,condition:creatingnextmemory,condition:coarsegrainingmemory,condition:publicannouncementsindicatememory} are satisfied. In particular, recall that the supports of the orthogonal projectors $\{\projmap_{\CP_j}^{(\effmem_j)}\}_{d\in\effmemSet_{j-1}}$ determines the possible announcements for each memory effect. Let the $A_j$ systems be a trivial system, which can be omitted from all expressions, and let $\QKDGmap{j} \in \CPTP(B_j \Mem_{j-1}, S_j \CP_j \CPhat_j \Mem_{j})$.

For each value of $d$ that can be stored in $\EffMem_j$, let $f^{(\effmem_j)}$ be a tradeoff function defined on the support of $\projmap_{\CP_j}^{(\effmem_j)}$, and let 
\begin{equation}
\label{eq:unconstrained-branch-kappa}
\begin{aligned}
\kappafuncgeneric{f^{(\effmem_j)}}{\mathord\cdot}{\QKDGmap{j}^{(\effmem_j)}}{\attack{j}} &\defvar
\inf_{\nu\in\Sigma_j^{(\effmem_j)}(\attack{j})}
\Halpha[f^{(\effmem_j)}]
(S_j|\CP_jE_j\widehat E)_\nu. \\
\Sigma_j^{(\effmem_j)}(\attack{j})
&\defvar
\left\{
\QKDGmap{j}^{(\effmem_j)}\circ\attack{j}
[\omega_{E_{j-1}\widehat E}]
\;\middle|\;
\omega\in\dop{=}(E_{j-1}\widehat E)
\right\},
\end{aligned}
\end{equation}
where $\widehat{E}$ is a purifying system, and $\alpha\in(1,\infty)$. 
Define a tradeoff function on $\CP_j$ as
\begin{equation}
\label{eq:branchwise-normalized-tradeoff}
\begin{aligned}
\fhat(\cP_j)
\defvar{}&f^{(\effmem_j)}_{|\cP_1^{j-1}}(\cP_j)
+\kappafuncgeneric{f^{(\effmem_j)}_{|\cP_1^{j-1}}}{\mathord\cdot}{\QKDGmap{j}^{(\effmem_j)}}{\attack{j}} \qquad \text{when $\cP_j \in \mathrm{supp}(\projmap_{\CP_j}^{(\effmem_j)})$ }
\end{aligned}
\end{equation}
Then the single-round optimization  $  \kappafuncgeneric{\fhat}{\sigmaconstraint{j}}{\QKDGmap{j}}{\attack{j}} $ that would be obtained through a direct application of the MEAT for this tradeoff function is greater than zero. That is, 
\begin{equation} \label{eq:historykappadefined_second}
\begin{aligned}
    \kappafuncgeneric{\fhat}{\cdot}{\QKDGmap{j}}{\attack{j}} &\defvar \inf_{\nu\in\Sigma_j^{\mathrm{hist}}(\attack{j})}
\Halpha[\fhat](S_j|\CP_j E_j\Mem_j\widetilde E)_\nu \geq 0,  \\
\Sigma_j^{\mathrm{hist}}(\attack{j})
&\defvar\bigl\{\,
\QKDGmap{j}\circ
(\attack{j}\otimes\id_{\Mem_{j-1}})
[\omega_{E_{j-1}\Mem_{j-1}\widetilde E}]
\;\big|\;
\omega\in\dop{=}(E_{j-1}\Mem_{j-1}\widetilde E) \big\}
\end{aligned}
\end{equation}
where $\widetilde{E}$ is a purifying system.
\end{lemma}
\begin{proof}
First, we repeat the proof of \cref{lem:effective-history-reduction}. This allows us to replace our goal (\cref{eq:historykappadefined_second}) with 
\begin{equation} \label{eq:effectivekappadefined_second}
\begin{aligned}
     \kappafuncgeneric{\fhat}{\cdot}{\effectiveQKDGmap{j}}{\attack{j}} & = \inf_{\nu\in\Sigma_j^{\mathrm{eff}}(\attack{j})}
\Halpha[\fhat](S_j|\CP_jE_j\widehat{E}')_\nu \geq 0,  \\
\Sigma_j^{\mathrm{eff}}(\attack{j})
&\defvar\bigl\{\,
\effectiveQKDGmap{j}\circ
(\attack{j}\otimes\id_{\EffMem_{j}})
[\omega_{E_{j-1}\EffMem_{j}\widehat{E}'}]
\;\big|\;
\omega\in\dop{=}(E_{j-1}\EffMem_{j}\widehat{E}')\},
\end{aligned}
\end{equation}
where $\widehat{E}'$ is a purifying system. Recall from \cref{condition:coarsegrainingmemory} that we have $\effectiveQKDGmap{j} = \Tr_{\EffMem_j} \circ \left( \sum_{\effmem_j} \ketbra{\effmem_j}_{\EffMem_j} \otimes \QKDGmap{j}^{(\effmem_j)}   \right)$. Thus, we obtain
\begin{equation}
     \Halpha[\fhat](S_j|\CP_jE_j\widehat{E}')_{\effectiveQKDGmap{j} \circ \attack{j} [    \omega_{ E_{j-1} \EffMem_j \widehat{E}'} ]}  \geq   \Halpha[\fhat](S_j|\CP_j \EffMem_j E_j\widehat{E}')_{\left( \sum_{\effmem_j} \ketbra{\effmem_j} \otimes \QKDGmap{j}^{(\effmem_j)} \right) \circ \attack{j} [    \omega_{ E_{j-1} \EffMem_j \widehat{E}'} ]}
\end{equation}
via data processing (\cref{lemma:DPIfweighted}). It is enough to show that RHS of the above inequality is greater than $0$, which (using \cref{def:fweighted}) is equivalent to 
\begin{equation}
    \sum_{\cP_j} \Pr(\Omega(\cP_j))_{\nu} 2^{ \left( \frac{1-\alpha}{\alpha} \right) \left(  \Halpha(S_j | \EffMem_j E_j \widehat{E}')_{\nu_{| \Omega(\cP_j)}} - \fhat(\cP)   \right)} \leq 1
\end{equation}
where $\nu_{S_j \CP_j \EffMem_j E_j \widehat{E}'} = \left( \sum_{\effmem_j} \ketbra{\effmem_j}_{\EffMem_j} \otimes \QKDGmap{j}^{(\effmem_j)}  \right) \circ \attack{j} [\omega_{ E_{j-1} \EffMem_j \widehat{E}'}]$, and $\Omega(\cP_j)$ is the event that $\CP_j$ stores $\cP_j$. 
To prove the above inequality, consider any state $\omega_{  E_{j-1} \EffMem_j \widehat{E}'} \in \Sigma_j^{\mathrm{eff}}(\attack{j})$ (which is, without loss of generality, classical in $\EffMem_j$), which we can write as 
\begin{equation}
    \omega_{ E_{j-1} \EffMem_j \widehat{E}'} = \sum_{\effmem_j} \Pr(\Omega(\effmem_j))_\omega \ketbra{\effmem_j}_{\EffMem_j} \otimes     \omega_{ E_{j-1}  \widehat{E}' | \Omega(\effmem_j)}.
\end{equation}
Moreover the event $\Omega(\effmem_j)$ is ``preserved" throughout, since $\left( \sum_{\effmem_j} \ketbra{\effmem_j}_{\EffMem_j} \otimes \QKDGmap{j}^{(\effmem_j)}  \right) \circ \attack{j}$ does not disturb the $\EffMem_j$ register.
Thus, we obtain 
\begin{align}
  &   \sum_{\cP_j} \Pr(\Omega(\cP_j))_{\nu} 2^{\left( \frac{1-\alpha}{\alpha} \right) \left(  \Halpha(S_j | \EffMem_j E_j \widehat{E}')_{\nu_{| \Omega(\cP_j)}} - \fhat(\cP)   \right)}      \nonumber  \\
 &= \sum_{\cP_j} \Pr(\Omega(\cP_j))_{\nu}  \sum_{\effmem_j} \Pr(\Omega(\effmem_j) | \Omega(\cP_j))  2^{\left( \frac{1-\alpha}{\alpha} \right) \left(  \Halpha(S_j | \EffMem_j E_j \widehat{E}')_{\nu_{| \Omega(\cP_j) \wedge \Omega(\effmem_j)}} - \fhat(\cP)   \right)}    \label{eq:seplemmatemp2}  \\
 & = \sum_{\effmem_j} \Pr(\Omega(\effmem_j))_\nu   \sum_{\cP_j} \Pr(\Omega(\cP_j) | \Omega(\effmem_j) )_{\nu}  2^{\left( \frac{1-\alpha}{\alpha} \right) \left(  \Halpha(S_j | \EffMem_j E_j \widehat{E}')_{\nu_{| \Omega(\cP_j) \wedge \Omega(\effmem_j)}} - \fhat(\cP)   \right)}  \label{eq:seplemmatemp3}   \\
  & = \sum_{\effmem_j} \Pr(\Omega(\effmem_j))_\nu   \sum_{\cP_j \in \mathrm{supp}\left(\projmap_{\CP_j}^{(\effmem_j)} \right) } \Pr(\Omega(\cP_j) | \Omega(\effmem_j) )_{\nu}  2^{\left( \frac{1-\alpha}{\alpha} \right) \left(  \Halpha(S_j | \EffMem_j E_j \widehat{E}')_{\nu_{| \Omega(\cP_j) \wedge \Omega(\effmem_j)}} - f^{(\effmem_j)}(\cP)   - \kappafuncgeneric{f^{(\effmem_j)}_{|\cP_1^{j-1}}}{\mathord\cdot}{\QKDGmap{j}^{(\effmem_j)}}{\attack{j}} \right)}  \label{eq:seplemmatemp4}   \\
  &\leq  \sum_{\effmem_j} \Pr(\Omega(\effmem_j))_\nu   = 1 \label{eq:seplemmatemp5}.
    \end{align}
Here, \cref{eq:seplemmatemp2} follows from the standard property of {\Renyi} entropies \cite[Proposition 5.1]{tomamichel_quantum_2016}\footnote{This can also be seen as setting $f=0$ in \cref{def:fweighted}}. In \cref{eq:seplemmatemp3} we use Bayes' rule. In \cref{eq:seplemmatemp4} we use the fact that \cref{condition:publicannouncementsindicatememory} allow us to conclude that conditioned on $\Omega(\effmem_j)$, only certain announcements can take place. We also used the definition of $\fhat$ in terms of $f^{(\effmem_j)}$ and $\kappafuncgeneric{f^{(\effmem_j)}_{|\cP_1^{j-1}}}{\mathord\cdot}{\QKDGmap{j}^{(\effmem_j)}}{\attack{j}}$.  For \cref{eq:seplemmatemp5} we use the fact that every inner sum (over $\cP_j \in \mathrm{supp}(\projmap_{\CP_j}^{(\effmem_j)})$) is less than $1$, which in turn follows from the definition of $\kappafuncgeneric{f^{(\effmem_j)}_{|\cP_1^{j-1}}}{\mathord\cdot}{\QKDGmap{j}^{(\effmem_j)}}{\attack{j}}$ from \cref{eq:unconstrained-branch-kappa}.  This concludes the proof. 
\end{proof}

If we use the above lemma together with MEAT, we obtain the following statement.

\begin{theorem}[MEAT for detector correlations with separate optimizations without marginal constraints]
\label{thm:meat-effective-history-nomarginal} 
Consider the scenario in \cref{fig:meat_attack_memory,lem:effective-history-reduction_nomarginal} , and suppose \cref{condition:classicalmemoryandmap,condition:creatingnextmemory,condition:coarsegrainingmemory,condition:publicannouncementsindicatememory} are satisfied. In particular, recall that the supports of the orthogonal projectors $\{\projmap_{\CP_j}^{(\effmem_j)}\}_{\effmem_j \in\effmemSet_{j}}$ determines the possible announcements for each memory effect. Let the $A_j$ systems be trivial systems, and let the sequence of channels be given by $\rho_{S_1^n \CP_1^n M_n E_n} = \QKDGmap{n} \circ \attack{n} \dots  \QKDGmap{1} \circ \attack{1}  [\sourcesymbol_{E_0}]$, where we have an arbitrary input state $\sigma_{E_0}$, and have $\QKDGmap{j} \in \CPTP(B_j \Mem_{j-1}, S_j \CP_j \CPhat_j \Mem_j)$.\footnote{When applied to entanglement-based QKD, the $B_j$ system will now include the registers sent by Eve to both Alice and Bob.}

For every value $j$, and every value of $\cP_1^{j-1}$, let $f^{(\effmem_j)}_{|\cP_1^{j-1}}$ be any tradeoff function on the support of $\projmap_{\CP_j}^{(\effmem_j)}$. Define a normalized tradeoff function on $\CP_j$ as
\begin{equation}
\begin{aligned}
\fhat_{| \cP_1^{j-1}}(\cP_j)
\defvar{}&f^{(\effmem_j)}_{|\cP_1^{j-1}}(\cP_j)
+\kappafuncgeneric{f^{(\effmem_j)}_{|\cP_1^{j-1}}}{\mathord\cdot}{\QKDGmap{j}^{(\effmem_j)}}{\attack{j}} \qquad \text{when $\cP_j \in \mathrm{supp}(\projmap_{\CP_j}^{(\effmem_j)})$ },
\end{aligned}
\end{equation}
where $\kappafuncgeneric{f^{(\effmem_j)}_{|\cP_1^{j-1}}}{\mathord\cdot}{\QKDGmap{j}^{(\effmem_j)}}{\attack{j}}$ is given by \cref{eq:unconstrained-branch-kappa}.
Define the global tradeoff function as $
\fhatfull(\cP_1^n)
\defvar
\sum_{j=1}^n\fhat_{|\cP_1^{j-1}}(\cP_j)$, and let $\alpha \in (1,\infty)$. Then, 
\begin{equation}
    \Halpha[\fhatfull](S_1^n|\CP_1^nE_n)_\rho
    \geq
    \Halpha[\fhatfull](S_1^n|\CP_1^nE_n\Mem_n)_\rho
    \geq0.
\end{equation}
\end{theorem}

\begin{proof}
The proof follows in two simple steps. First, we use \cref{thm:meatqkd}, which is simply the direct application of the MEAT \cite{arqand_marginal_2025} for the setting considered in \cref{fig:meat_attack_memory}, with $\Mem_{j-1} E_{j-1}$ as Eve's side-information register. The second step is the application of \cref{lem:effective-history-reduction_nomarginal} which replaces the single-round optimization. 
    \end{proof}

\subsection{Discussion and Examples}
\label{subsec:detectorexamples}
We will now describe how certain correlated detector imperfections can be analyzed with \cref{thm:meat-effective-history,thm:meat-effective-history-nomarginal}. When analyzing prepare-and-measure protocols, we should use \cref{thm:meat-effective-history}, since it preserves the marginal constraint. For entanglement based protocols, one can use either theorem: however, \cref{thm:meat-effective-history-nomarginal} offers cheaper numerics and is the better choice.

\subsubsection{Dead times and afterpulsing} \label{subsubsec:deadtimesandafterpulsing}
Let us consider a scenario where detectors have dead times and exhibit afterpulsing. We assume that these effects last for some finite length $\lcdet$ of rounds. That is, for any given round, the detectors exhibit correlated behaviour if and only if there were (some) clicks in any the previous $\lcdet$ rounds. Otherwise, the detector behaviour is independent of previous rounds. To analyze this scenario using this work, we are required to make the following protocol modification: We consider a protocol where after a click is obtained in a given round, the detectors are shut off (via suitable hardware modifications) for the next $\lcdet$ rounds and therefore do not register any clicks. Moreover, during these $\lcdet$ rounds where detectors are shut off, Bob announces the special symbol $\discard$ in the public announcement register $\CP_j$, and Alice stores $\bot$ in $S_j$. Note that these restrictions are identical to those from Refs.~\cite{nahar2025imperfect,wang2025phase}, and still represent the most general scenario of detector correlations that can be handled in QKD proofs.

This can be modeled in \cref{fig:meat_attack_memory} as follows. We let $\Mem_j$ store the exact information about which of the previous rounds registered a click, or did not register a click.\footnote{Technically, it suffices if we only let $\Mem_j$ store the earliest click.  However, since the dimensions of $\Mem_j$ never shows up in the numerical computations we perform, we do not make any attempts to reduce its dimensions. } Thus, for a correlation length of $\lcdet$, $\Mem_{j-1}$ stores $\lcdet$-bit strings. The $\coarsegrainedhistory{j}$ map reads $\Mem_{j-1}$ and creates the $\EffMem_j$ register which stores a binary value $\effmem_j$, indicating whether there was a click in the previous $\lcdet$ rounds. This is then followed by the $\effectiveQKDGmap{j}$ map, which applies the map $\QKDGmap{j}^{(\effmem_j)}$ depending on the value stored in the $\EffMem_j$ register. If $\effmem_j$ corresponds to no clicks in the previous $\lcdet$ rounds, then $\QKDGmap{j}^{(\effmem_j)}$ is simply the ``usual" map: Alice and Bob measure using some known, fixed POVM, and undertake their usual announcements and postprocessing.
If on the other hand, $\effmem_j$ corresponds to a click in the previous $\lcdet$ rounds, then $\QKDGmap{j}^{(\effmem_j)}$ measures using a trivial POVM with a single outcome, stores $\bot$ in $S_j$, and announces the special symbol $\discard$ in $\CP_j,\CPhat_j$. Finally, the new memory register $M_j$ can be deterministically constructed from $\Mem_{j-1}$ and $\CP_j$.  Thus, \cref{condition:classicalmemoryandmap,condition:coarsegrainingmemory,condition:creatingnextmemory,condition:publicannouncementsindicatememory} are all satisfied, and \cref{thm:meat-effective-history,thm:meat-effective-history-nomarginal} can be applied.  We compute key rates utilizing this approach in \cref{sec:plots}. 

If we use \cref{thm:meat-effective-history}, then $\EffMem_j$ is an additional $2$-dimensional register, and we add two extra announcements in the $\CP_j$ register  corresponding to $\discard$ for test and generation rounds. Thus, there is a small increase in numerical complexity, which is independent of the strength or length of the correlated effects. 
If we use \cref{thm:meat-effective-history-nomarginal} (for entanglement-based protocols), then we can simply choose the tradeoff function $\fhat^{(\effmem_j)}$ corresponding to there being clicks in the previous $\lcdet$ rounds, to always output zero. This allows to satisfy the required property (see \cref{lem:effective-history-reduction_nomarginal}) by construction. In this case, we only have to evaluate the sole remaining $\fhat^{(\effmem_j)}$, and there is no added numerical complexity. 

\begin{remark}
At this stage, one may be wonder if a stronger result can be obtained. Our results require discarding of rounds, and shutting off detectors, even if the strength of the correlated imperfections is very weak (which is indeed the case for afterpulsing).  While we believe such a result should be possible, it still remains an interesting open problem. We stress that it is \emph{not} resolved by secret memory registers that require the no-signaling condition (see \cref{remark:conditioning_imperfections}). In this work, we are unable to allow $\QKDGmap{j}$ to depend on the exact outcomes of previous round  measurements, since doing so requires $\mathcal{M}_j$ to receive the click pattern of the previous rounds, which also leaks it to Eve. For the same reason, we forced to physically shut off the detector. Note that from a performance perspective, it is always desirable to shut off detectors for any round that we know we are going to discard anyway, since a click in that round may cause us to further discard future rounds. Fortunately, we note that such hardware modifications are certainly possible in current experiments \cite{Makarov_preparingacommericial_2024}. 
\end{remark}

Finally, one may wish to analyze detectors with unbounded memory
effects. A possible approach is to introduce a finite cutoff
$\lcdet$, and use the approach described earlier. In this case, one must rigorously account for the deviation introduced by neglecting memory effects beyond the chosen cutoff length. 
To obtain a security guarantee for the actual
detectors, one must additionally bound the trace distance between
the final protocol states generated by the actual and approximate
models. This can indeed be done using \cite[Theorem 7.1]{nahar_phd_2025}, as long as the dependence of the POVM of the later rounds on the outcome of previous rounds is weak.

\subsubsection{Iterative sifting for entanglement-based protocols}
Our results directly allow us to accommodate iterative sifting, for entanglement-based protocols, with any stopping
condition that is a deterministic function of the public announcements. Iterative sifting refers to the scenario where the total number of rounds in the protocol are not fixed, but determined by the public announcements during the protocol. For instance, one may wish to implement a protocol where Alice and Bob continue until a fixed number of detections are observed. The only work that explicitly analyzes iterative sifting is Ref.~\cite{tamaki_security_2018}, which can accommodate certain kinds of iterative sifting, for ideal hardware. Note that on-the-fly announcements are necessary to implement iterative sifting. We model iterative sifting for entanglement-based protocols in \cref{fig:meat_attack_memory} as follows. (We comment on the difficulty in extending this to prepare-and-measure protocols shortly.)

We assume that the protocol can have 
maximum number \(n\) of rounds and apply our theorems to a fixed
sequence of \(n\) channels. Only some of these channels actually correspond to operations in the actual protocol. 
Let $
    f_\mathrm{stop}(j,\cP_1^{j-1})
$
be a binary-valued function of the round index \(j\) and all preceding
announcements, indicating whether the protocol should stop or
continue. 
We let the register \(\Mem_{j-1}\) store a copy of the preceding public
announcements \(\CP_1^{j-1}\). The map \(\coarsegrainedhistory{j}\)
reads this register and creates \(\EffMem_j\), which stores the value
\(\effmem_j = f_\mathrm{stop}(j,\cP_1^{j-1})\). The map
\(\effectiveQKDGmap{j}\) then applies \(\QKDGmap{j}^{(\effmem_j)}\).
If \(\effmem_j\) is \texttt{continue}, this is the usual map: Alice and Bob
measure using a known, fixed POVM and make their usual announcements.
If \(\effmem_j\) is \texttt{stop}, the map instead uses a trivial POVM with
a single outcome, stores \(\bot\) in \(S_j\), and announces
\(\stopannouncement\) in \(\CP_j\). This corresponds to the scenario where the protocol has stopped, and these channels only occur in the theoretical analysis of the protocol. Finally,
\(\historyupdatemap{j}\) creates \(\Mem_j\) by appending a copy of
\(\CP_j\) to \(\Mem_{j-1}\). Note that defining \(f_\mathrm{stop}(j,\cP_1^{j-1})\) to output \texttt{stop}
whenever \(\cP_1^{j-1}\) contains \(\stopannouncement\) ensures that
all rounds following a stop announcement are discarded. 

Now, due to \cref{thm:meat-effective-history-nomarginal}, we need to compute two tradeoff functions, one for the case where $\EffMem_j$ stores \texttt{stop} and the other for when $\EffMem_j$ stores \texttt{continue}. The former can be straightforwardly chosen to always output zero, while the latter can be computed using standard methods, since it corresponds to the case where there are no memory effects. Then, \cref{thm:meat-effective-history-nomarginal} allows us to combine the two into the tradeoff function we implement in the protocol. This allows us to accommodates a very wide class of stopping conditions.\footnote{In fact, we can handle some very general scenarios where, in each round, Eve can simply ``tell" Alice and Bob whether they should utilize that round normally, or discard it. Intuitively, for entanglement-based protocols, Eve anyway chooses the entirety of the states that Alice and Bob measure, and therefore allowing her to decide which states are chosen gives her no additional power. (If she wanted Alice and Bob to discard a round, she would not have sent that state to them in the first place.) } For some concrete
examples, consider the scenarios where, subject to the maximum number \(n\) of pulses, Alice and Bob
 continue until:
\begin{enumerate}
    \item The total number of detected outcomes reaches a prescribed
    threshold, or
    
    \item The number of detected outcomes in key generation rounds
    reaches a prescribed threshold, or
    
    \item A publicly computable tradeoff function output for the rounds completed
    so far reaches a prescribed threshold. Such a stopping condition could be used to gather
    enough data to produce a key of a prescribed length, provided that
    the threshold also accounts for the other terms in the key-length
    bound, including error-correction leakage and finite-size
    corrections.
\end{enumerate}
In each case, the stopping decision must be determined by the public
announcements available at that point in the protocol.

Notice that in the above analysis we used \cref{thm:meat-effective-history-nomarginal}, since it allows us to build the full tradeoff functions from the analysis of each memory effect separately. In the case of prepare-and-measure protocols, we cannot do this. In that case, one may attempt to solve the full optimization for a given round using \cref{thm:meat-effective-history}. Here,  one runs into the problem described in point (3) at the start of this section; namely, that one needs to perform optimizations overall possible states $\omega_{A_j M_{j-1} E_{j-1} \widetilde{E}}$, which contains states where $M_{j-1}$ always corresponds to the protocol stopping. We could avoid this problem for detector correlations by announcing, in the test rounds, whether the detectors were turned on or off; however, the same option is not available in the case of iterative sifting (since after the protocol stops, there is no notion of test or generation rounds , and in any case, the stopping of the protocol is a single discrete event and not something that occurs repeatedly). We leave an analysis of iterative sifting for prepare-and-measure protocols to future work.

\section{MEAT for source correlations}\label{sec:source-correlations}

We now turn to QKD protocols whose sources exhibit correlated
behaviour. As we explain in \cref{subsec:sourcereplacedstates}, the
corresponding source-replaced states need not have a tensor-product
structure on the \(A_j\) systems, so the MEAT cannot be applied
directly. We assume independent setting choices and a finite correlation length
\(\lcsrc\), where the state prepared in
each round depends only on the current setting and the preceding
\(\lcsrc\) settings. Following the approach of
Ref.~\cite{curraslorenzo_phaseerrorcorrelated_2026}, we analyze several
interleaved subsequences of the protocol. For intuition, consider \(\lcsrc=1\), and refer to \cref{fig:Meatsourcecorrelations}. The argument proceeds in four steps:
\begin{enumerate}
    \item We measure Alice's setting registers ($\Ameas$) in all even-numbered
    rounds and record their outcomes.

    \item Conditioned on an outcome string \(u\) for these registers,
    we recover the required tensor-product structure of the source-replaced state by grouping
    rounds into pairs \((1,2),(3,4),\ldots\) . This is the central observation from Ref.~\cite{curraslorenzo_phaseerrorcorrelated_2026}  

    \item We group the corresponding channels together and
    apply the MEAT to bound the entropy of the odd-numbered secret
    registers.

    \item We repeat the argument with the roles of odd and even rounds
    reversed to bound the entropy of the even-numbered secret
    registers. We then combine the two bounds using the weighted
    chain rule in \cref{lem:weighted-two-block-chain-rule}.
\end{enumerate}

For the general case, we require partitioning into \(\lcsrc+1\)
interleaved subsequences. To analyze one subsequence, we condition on
the settings of all the other sequences and group the physical rounds so that the
MEAT can be applied. Repeating this procedure
for all \(\lcsrc+1\) subsequences and combining the resulting entropy
bounds gives a bound for the full protocol. The argument follows the
same steps as the case \(\lcsrc=1\), but requires additional notation
to describe the interleaving and grouping. We introduce this notation
first.
\subsection{Notation}
For a given correlation length $\lcsrc$, we will need to consider 
 $\corrperiod\defvar\lcsrc+1$ interleaved protocols. For each $r\in[\corrperiod]$, we define the set of selected rounds (whose entropy we wish to bound) and their complement (whose setting choices are measured) by
\begin{equation}
 \interleavedrounds{r}\defvar\{r+k\corrperiod:k\in\mathbb N_0\}\cap[n],
 \qquad
 \otherinterleavedrounds{r}\defvar[n]\setminus\interleavedrounds{r}.
 \label{eq:source-interleaved-rounds}
\end{equation}
We define $m_r\defvar|\interleavedrounds{r}|$. For \(k\in[m_r]\),  we define the \(k\)-th selected round index by
\begin{equation}
    \selectedround{k}{r} \defvar r+(k-1)\corrperiod.
\end{equation}
We will often write $\selectedround{k}{r}$ as $j$ for brevity. 
The associated blocks are given by
\begin{equation}
\begin{aligned}
    \sourceblock{0}{r}
        &\defvar [\min\{r-1,n\}], \\
    \sourceblock{k}{r}
        &\defvar
        \left\{
            \selectedround{k}{r},\ldots,
            \min\{\selectedround{k}{r}+\lcsrc,n\}
        \right\},
        \qquad k\in[m_r], \\
    \sourceblock{k}{r}'
        &\defvar
        \sourceblock{k}{r}\setminus\{\selectedround{k}{r}\},
        \qquad k\in[m_r].
\end{aligned}
\end{equation}
Thus, \(\sourceblock{0}{r}\) contains the rounds preceding the first
selected round, while each subsequent block contains a selected round
and the following \(\lcsrc\) rounds, truncated at \(n\).
The primed block $\sourceblock{k}{r}'$ excludes the selected round.
Notice that each block with $k\geq1$ starts with a selected round and contains no other selected round. 

For $V\subseteq[n]$, we use $X_V$ to denote the collection of registers $X_q$ with $q\in V$.  For \(1\leq i\leq j\leq n\), we use
\begin{equation}
    X_{i,r}^{j}
        \defvar X_{\{i,i+1,\ldots,j\}\cap\interleavedrounds{r} },
    \qquad
    X_{i,\neq r}^{j}
        \defvar X_{\{i,i+1,\ldots,j\}\cap\otherinterleavedrounds{r}  }.
\end{equation}
 Thus,
\(X_{i,r}^{j}\) denotes the registers between rounds \(i\) and \(j\)
that belong to the \(r\)th interleaved subsequence, while
\(X_{i,\neq r}^{j}\) denotes those that belong to the other
subsequences. In particular, we have $
    X_{1,r}^{n}=X_{\interleavedrounds{r}},
    X_{1,\neq r}^{n}=X_{\otherinterleavedrounds{r}}$.

As outlined earlier, our proof involves preemptively measuring the
\(\Ameas_j\) registers for \(j\in\otherinterleavedrounds{r}\).
We define \(\measmap{j}\in\CPTP(A_j,A_j)\), where
\(A_j=\Ameas_j\Ashield_j\), to be the channel that measures
\(\Ameas_j\) in the setting basis, retains the outcome in the same
register, and leaves \(\Ashield_j\) untouched. For
\(V\subseteq[n]\), we write $
    \measmap{V}\defvar\bigotimes_{j\in V}\measmap{j} $
for the corresponding measurement channel on the rounds in \(V\).

\begin{condition}[Protocol maps measure \(\Ameas_j\)]
\label{condition:protocolmapsmeasAj}
To justify that these preemptive measurements leave the final state
unchanged, we require each protocol map to satisfy
\begin{equation}
    \QKDGmap{j}
        = \QKDGmap{j}\circ\measmap{j}.
    \label{eq:pinchingdoesnotchangeGmap}
\end{equation}
This condition holds whenever the first step of \(\QKDGmap{j}\)
measures \(\Ameas_j\) in the setting basis, which is the case for the
source-replacement description of the QKD protocols considered here.
\end{condition}

\subsection{Source-replaced states with finite classical correlation length}\label{subsec:sourcereplacedstates}

For an independent source, Alice chooses $i_j\in[d_A]$ with probability $p^{(j)}_{i_j}$ and prepares the signal state $ \left( \sigma^{(j)}_{i_j}\right)_{A'_j}$. The source-replaced state is then given by 
\begin{align}
 \ket{\psi^{(j)}}_{\Ameas_j\Ashield_jA'_j}
 &\defvar\sum_{i_j=1}^{d_A}\sqrt{p^{(j)}_{i_j}}
 \ket{i_j}_{\Ameas_j}\ket{\sigma^{(j)}_{i_j}}_{\Ashield_jA'_j},\\
 \sourcesymbol^{\mathrm{ind}}_{A_1^n(A')_1^n}
 &\defvar\bigotimes_{j=1}^n\ketbra{\psi^{(j)}}_{A_jA'_j}.
 \label{eq:independent-source-replaced}
\end{align}
where $\ket{\sigma^{(j)}_{i_j}}_{\Ashield_jA'_j}$ is a purification of $\left( \sigma^{(j)}_{i_j}\right)_{A'_j}$. 
Its marginal on $A_1^n$ is a tensor product, as required in \cref{thm:meatqkd}. We will now formulate the required properties of the correlated state preparation to which our results are applicable.

\begin{condition} \label{condition:sourcemodel}
We consider a correlated source where the setting choices in each round are independent, but the state emitted in round $j$ may depend on the setting choices of the previous $\lcsrc$ rounds. Thus, in round $j$, Alice prepares $ \left(\sigma^{(j,i_{j-\lcsrc}^{j-1})}_{i_j} \right)_{A'_j}$, where $i_j$ denotes her choice on the $j$th round, and $i_{j-\lcsrc}^{j-1}$ denotes the list of setting choices in the previous $\lcsrc$ rounds. 
Using $\ket{\sigma^{(j,i_{j-\lcsrc}^{j-1})}_{i_j}}_{\Ashield_jA'_j}$ to denote the purification of the state produced when Alice chooses $i_j$ in round $j$, the source-replaced state is given by
\begin{align}
 \ket{\Psi^{\mathrm{corr}}}_{A_1^n(A')_1^n}
 &\defvar\sum_{i_1^n}\left(\prod_{j=1}^n\sqrt{p^{(j)}_{i_j}}\right)
 \ket{i_1^n}_{\Ameas_1^n}
 \bigotimes_{j=1}^n\ket{\sigma^{(j,i_1^{j-1})}_{i_j}}_{\Ashield_jA'_j},\\
 \sourcesymbol^{\mathrm{corr}}_{A_1^n(A')_1^n}
 &\defvar\ketbra{\Psi^{\mathrm{corr}}}_{A_1^n(A')_1^n}.
 \label{eq:correlated-source-replaced}
\end{align}
 \end{condition}

It is straightforward to verify that the state $\sourcesymbol^{\mathrm{corr}}_{A_1^n}$ does not have the required tensor product structure on $A_1^n$. Our first task is to recover this structure.
The following pair of lemmas recovers tensor product structure after measuring $A_{\otherinterleavedrounds{r}}$ and conditioning on the recorded outcome. We split them up into two lemmas for readability, and provide a common proof.

\begin{lemma}[Recovering tensor-product structure on $\Ameas_1^n$.]
\label{lemma:measuretogettensorproduct}
Let $\sourcesymbol^{\mathrm{corr}}$ be the source state in
\cref{eq:correlated-source-replaced}, with independent settings and
correlation length $\lcsrc$ as specified in
\cref{condition:sourcemodel}. Fix $r\in[\corrperiod]$ and apply the
preemptive measurements map $\measmap{\otherinterleavedrounds{r}}$.
Let $u_{\otherinterleavedrounds{r}}$ denote the outcomes retained
in $\Ameas_{\otherinterleavedrounds{r}}$, and let
$\Omega(u_{\otherinterleavedrounds{r}})$ denote the event that
these registers have the specified values. 
Thus, we have
\begin{equation} \label{eq:measuredstateddefined}
\begin{aligned}
  \measuredsource{\otherinterleavedrounds{r}}
    _{A_1^n(A')_1^n}
   & \defvar
    \measmap{\otherinterleavedrounds{r}}
    [\sourcesymbol^{\mathrm{corr}}]. \\
    &=
    \sum_{u_{\otherinterleavedrounds{r}}}
    \Pr\bigl(\Omega(u_{\otherinterleavedrounds{r}})\bigr)\,
    \measuredsource{u_{\otherinterleavedrounds{r}}}
    _{A_1^n(A')_1^n},
    \\
    \measuredsource{u_{\otherinterleavedrounds{r}}}
    _{A_1^n(A')_1^n}
  &  \defvar
    \left(
        \measuredsource{\otherinterleavedrounds{r}}
        _{A_1^n(A')_1^n}
    \right)_{|\Omega(u_{\otherinterleavedrounds{r}})} \qquad\qquad \text{ (for brevity) } \\
    \Pr\bigl(\Omega(u_{\otherinterleavedrounds{r}})\bigr)
    &=
    \prod_{q\in\otherinterleavedrounds{r}}
    p^{(q)}_{u_q}.
\end{aligned}
\end{equation}
For \(k\in\{0,\ldots,m_r\}\), define the grouped registers $
    \blockversion{A}_{k,r}
        \defvar A_{\sourceblock{k}{r}}$, and 
        $    \blockversion{A'}_{k,r}
        \defvar A'_{\sourceblock{k}{r}}$
        . Then the states $ \measuredsource{u_{\otherinterleavedrounds{r}}}
    _{A_1^n(A')_1^n}$ are tensor product across $\blockversion{A}_{0,r}\blockversion{A}_{1,r}\cdots\blockversion{A}_{m_r,r}$, with its $k$th factor depending only on
$u_{\sourceblock{k-1}{r}'\cup\sourceblock{k}{r}'}$. That is, we have
\begin{equation}
    \measuredsource{u_{\otherinterleavedrounds{r}}}
    _{A_1^n(A')_1^n}
    =
    \measuredsource{u_{\sourceblock{0}{r}'}}
    _{\blockversion{A}_{0,r}\blockversion{A'}_{0,r}}
    \otimes \left( 
    \bigotimes_{k=1}^{m_r}
    \measuredsource{
        u_{\sourceblock{k-1}{r}'\cup\sourceblock{k}{r}'}
    }_{\blockversion{A}_{k,r}\blockversion{A'}_{k,r}} \right).
\label{eq:measuretogettensorproduct}
\end{equation}
\end{lemma}

\begin{lemma} \label{lemma:theblockstates}
Consider the same setting as \cref{lemma:measuretogettensorproduct}.   For $k\in[m_r]$, writing $j=\selectedround{k}{r}$, define
\begin{equation}
\begin{aligned}
    \ket{\psi^{(0,r)}_{u_{\sourceblock{0}{r}'}}}
    _{\blockversion{A}_{0,r}\blockversion{A'}_{0,r}}
    \defvar &
    \ket{u_{\sourceblock{0}{r}'}}
    _{\Ameas_{\sourceblock{0}{r}'}}
    \otimes \left( 
    \bigotimes_{q\in\sourceblock{0}{r}}
    \ket{\sigma^{(q,u_1^{q-1})}_{u_q}}
,    _{\Ashield_qA'_q} \right) \\
    \ket{
        \psi^{(k,r)}_{
            u_{\sourceblock{k-1}{r}'\cup\sourceblock{k}{r}'}
        }
    }_{\blockversion{A}_{k,r}\blockversion{A'}_{k,r}} \defvar &
    \sum_{i_j=1}^{d_A}
    \sqrt{p^{(j)}_{i_j}}\,
    \ket{i_j}_{\Ameas_j}
    \otimes
    \ket{u_{\sourceblock{k}{r}'}}
    _{\Ameas_{\sourceblock{k}{r}'}}
    \otimes \left( 
    \bigotimes_{q\in\sourceblock{k}{r}}
    \ket{\sigma^{(q,i_{q-\lcsrc}^{q-1})}_{i_q}}
    _{\Ashield_qA'_q} \right),
\end{aligned}
\label{eq:source-conditional-block-ket}
\end{equation}
where $i_q=u_q$ for $q\in\otherinterleavedrounds{r}$. Then the states on the right-hand side of \cref{eq:measuretogettensorproduct} are given
by 
\begin{align}
    \measuredsource{u_{\sourceblock{0}{r}'}}
    _{\blockversion{A}_{0,r}\blockversion{A'}_{0,r}}
    &\defvar
    \ketbra{\psi^{(0,r)}_{u_{\sourceblock{0}{r}'}}},
    \\
    \measuredsource{
        u_{\sourceblock{k-1}{r}'\cup\sourceblock{k}{r}'}
    }_{\blockversion{A}_{k,r}\blockversion{A'}_{k,r}}
    &\defvar
    \ketbra{
        \psi^{(k,r)}_{
            u_{\sourceblock{k-1}{r}'\cup\sourceblock{k}{r}'}
        }
    },
    \qquad k\in[m_r].
\end{align}
where the superscripts on $\sigma$ specify the entries of the
complete measurement outcome on which each block state depends.
\end{lemma}

\begin{proof} We prove both \cref{lemma:measuretogettensorproduct,lemma:theblockstates} together. 
Fix a measurement outcome
$u_{\otherinterleavedrounds{r}}$ of $\measmap{\otherinterleavedrounds{r}}$. Projecting the correlated source-replaced state from 
\cref{eq:correlated-source-replaced} onto this outcome gives
\begin{equation}
\ketbra{u_{\otherinterleavedrounds{r}}}    _{\Ameas_{\otherinterleavedrounds{r}}}
\ket{\Psi^{\mathrm{corr}}}_{A_1^n(A')_1^n}
=    \left(        \prod_{q\in\otherinterleavedrounds{r}}
        \sqrt{p^{(q)}_{u_q}}
    \right)
   \sum_{i_{\interleavedrounds{r}}}
    \left(      \prod_{j\in\interleavedrounds{r}}
        \sqrt{p^{(j)}_{i_j}}
    \right)
\ket{u_{\otherinterleavedrounds{r}}}    _{\Ameas_{\otherinterleavedrounds{r}}}
    \otimes
    \ket{i_{\interleavedrounds{r}}}
    _{\Ameas_{\interleavedrounds{r}}}
    \otimes \left( 
    \bigotimes_{q=1}^n
    \ket{\sigma^{(q,i_{q-\lcsrc}^{q-1})}_{i_q}}
    _{\Ashield_qA'_q} \right),
\label{eq:source-projected-ket}
\end{equation}
where $i_q=u_q$ for $q\in\otherinterleavedrounds{r}$.

Let $k\geq1$ and consider $\selectedround{k}{r}$, and the block $\sourceblock{k}{r}$ formed by that round with the next  $\lcsrc$ rounds. For each
$q\in\sourceblock{k}{r}$, the signal state for that round only depends on the previous indices 
   $ \{\max\{1,q-\lcsrc\},\ldots,q\} $, and thus contains $\selectedround{k}{r}$ (and no other selected round from $\interleavedrounds{r}$). Every other index
in this window belongs to
$\sourceblock{k-1}{r}'\cup\sourceblock{k}{r}'$, whose setting choices are fixed by our measurements on $\otherinterleavedrounds{r}$.
Consequently, we can rewrite
\cref{eq:source-projected-ket} and obtain
\begin{equation} \label{eq:tempone}
\ketbra{u_{\otherinterleavedrounds{r}}}  _{\Ameas_{\otherinterleavedrounds{r}}}\ket{\Psi^{\mathrm{corr}}}_{A_1^n(A')_1^n} =
    \left(        \prod_{q\in\otherinterleavedrounds{r}}
        \sqrt{p^{(q)}_{u_q}}
    \right)
\ket{\psi^{(0,r)}_{u_{\sourceblock{0}{r}'}}}
    \otimes
    \bigotimes_{k=1}^{m_r}
    \ket{
        \psi^{(k,r)}_{
            u_{\sourceblock{k-1}{r}'\cup\sourceblock{k}{r}'}
        }
    },
\end{equation}
where the states on the RHS of the above expression are as described in the lemma statement.

The fact that these pure states are normalized follows from
\begin{equation}
    \abs{
        \braket{
            \psi^{(k,r)}_{
                u_{\sourceblock{k-1}{r}'\cup\sourceblock{k}{r}'}
            }
        }
    }^2
    =
    \sum_{i_j=1}^{d_A}p^{(j)}_{i_j}
    =
    1.
\end{equation}
Furthermore, it is straightforward to verify that 

\begin{equation}
    \Pr\bigl(\Omega(u_{\otherinterleavedrounds{r}})\bigr)
    =
    \prod_{q\in\otherinterleavedrounds{r}}p^{(q)}_{u_q
    }.\end{equation}
Dividing the outer product of the pure state in \cref{eq:tempone} by this probability proves the desired result. 
\end{proof}

The above lemma allows us recover some tensor product structure on the $A_1^n$ systems. However, in doing so we had to measure some of the $A_j$ systems, and also condition on the specific outcome. Since the measurement we performed is exactly the same one performed by $\QKDGmap{j}$ (when \cref{condition:protocolmapsmeasAj} is satisfied), the output state in  \cref{fig:meat_attack} is unchanged as a result of these preemptive measurements. This is what allows us to freely perform these measurements without changing the scenario under consideration. This is stated formally in the following lemma\footnote{Note that equality of the final output states does not imply
that applying the MEAT before and after the preemptive measurements
gives the same entropy bound. The MEAT gives Eve a purification of
the marginal on the $A_j$ systems, and preemptive measurement changes
that marginal and hence the purification available to Eve in the
theorem's analysis. Thus, moving a measurement from inside the
protocol channels to the source state can change the resulting
bound, even though it leaves the actual protocol output unchanged.
The argument here only establishes that the both protocol output the same state, and does not claim that the application of the MEAT is equivalent in both cases.}:

\begin{lemma}[Preemptive measurements do not change  output state]\label{lem:early-setting-measurement} Consider the sequence of channels in \cref{fig:meat_attack}. 
For any $j \in [n]$, let $\measmap{j} \in \CPTP(A_j,A_j)$ be a channel such that $\QKDGmap{j} \circ \measmap{j} = \QKDGmap{j}$ (i.e, \cref{condition:protocolmapsmeasAj} is satisfied). Let $V \subseteq [n]$, and let $\measmap{V} = \otimes_{i \in V} \measmap{i}$. Then, 
\begin{equation}
    \begin{aligned}
        \QKDGmap{n} \circ \attack{n} \circ \dots \circ \QKDGmap{1} \circ \attack{1} =  \QKDGmap{n} \circ \attack{n} \circ \dots \circ \QKDGmap{1} \circ \attack{1}
    \circ \measmap{V}
    \end{aligned}
\end{equation}
\end{lemma}

\begin{proof}
The proof follows from the observation that $\measmap{i}$ commutes with $\attack{j}$ for all $j$, and commutes with $\QKDGmap{j}$ for all $j\neq i$. Thus we can move $\measmap{i}$ for any $i$ to the left in the sequence of channels until we reach the $\QKDGmap{i}$. Then, we can use \cref{condition:protocolmapsmeasAj}, which states that $\QKDGmap{i} \circ \measmap{i} = \QKDGmap{i}$ to absorb $\measmap{i}$. Since this can be done for any $i$, the required statement follows.
\end{proof}

\subsection{Proving the MEAT variant}
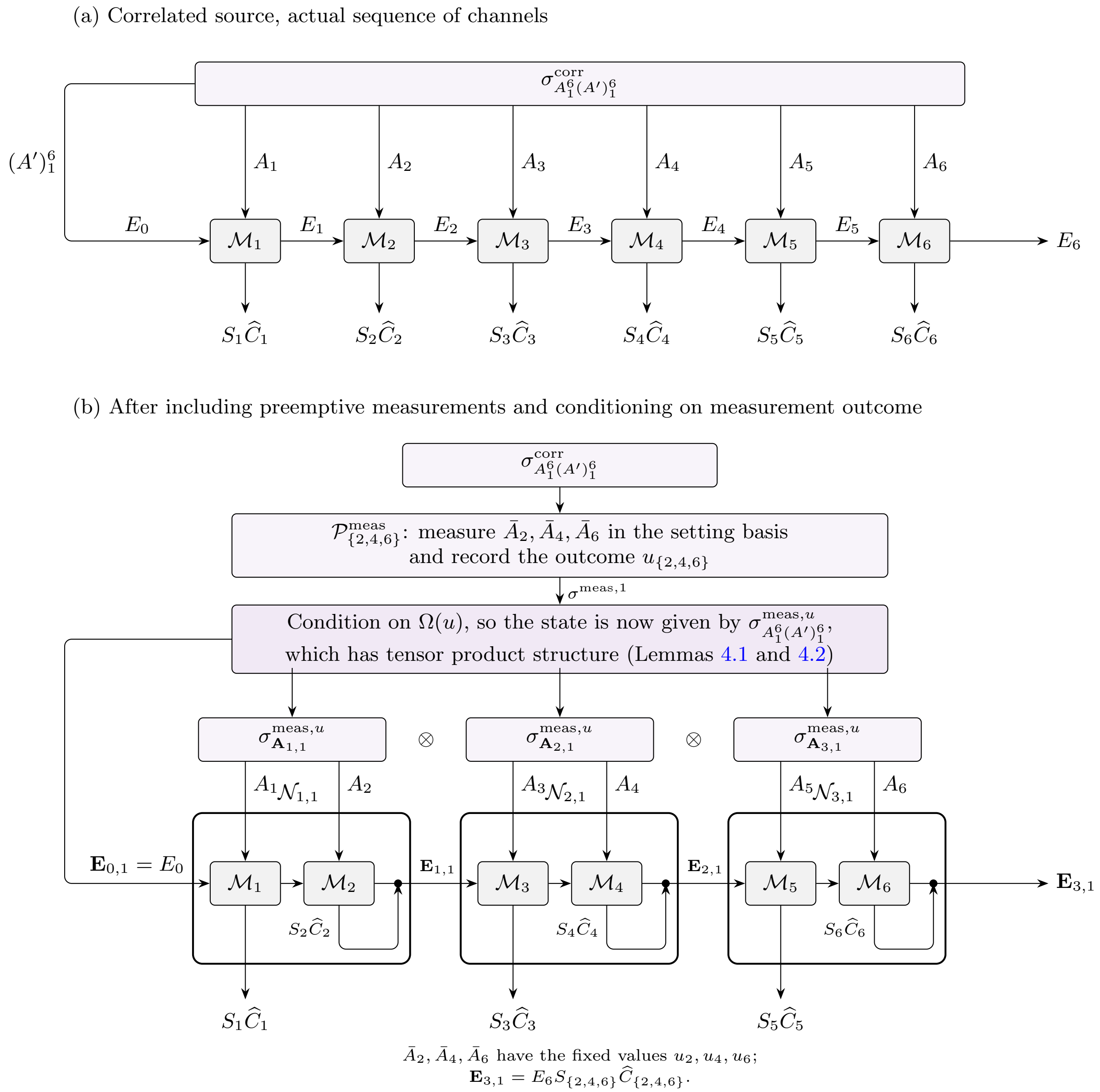
\begin{figure}[]
\centering
\begin{tikzpicture}[>=Stealth,font=\small,
 roundbox/.style={draw,rounded corners=2pt,minimum width=1.05cm,minimum height=.65cm,fill=gray!10},
 sourcebox/.style={draw,rounded corners=2pt,minimum height=.65cm,align=center,fill=llmrevision!5},
 measurebox/.style={sourcebox,minimum width=9.8cm,minimum height=.85cm},
 signal/.style={->,rounded corners=3pt},
 junction/.style={circle,fill,inner sep=1.2pt}]
\node[anchor=west] at (-2.7,3.35) {(a) Correlated source, actual sequence of channels};
\node[sourcebox,minimum width=11.5cm] (source) at (5,2.35)
 {$\sourcesymbol^{\mathrm{corr}}_{A_1^6(A')_1^6}$};
\foreach \q/\x in {1/0,2/2,3/4,4/6,5/8,6/10}{
 \node[roundbox] (m\q) at (\x,0) {$\onerndchann_{\q}$};
 \draw[signal] (\x,2.02) -- node[right] {$A_{\q}$} (m\q.north);
 \draw[signal] (m\q.south) -- ++(0,-.75) node[below] {$S_{\q}\CP_{\q}$};
}
\draw[signal] (source.west) -- (-2.7,2.35)
 -- node[left] {$(A')_1^6$} (-2.7,0)
 -- node[above] {$E_0$} (m1.west);
\foreach \q/\next in {1/2,2/3,3/4,4/5,5/6}{
 \draw[signal] (m\q.east) -- node[above] {$E_{\q}$} (m\next.west);
}
\draw[signal] (m6.east) -- (12,0) node[right] {$E_6$};

\begin{scope}[yshift=-8.4cm]
\node[anchor=west] at (-2.7,5.9) {(b) After including preemptive measurements and conditioning on measurement outcome};
\node[sourcebox,minimum width=4.7cm] (corr) at (4.7,5.05)
 {$\sourcesymbol^{\mathrm{corr}}_{A_1^6(A')_1^6}$};
\node[measurebox] (measure) at (4.7,3.85)
 {$\measmap{\{2,4,6\}}$: measure $\Ameas_2,\Ameas_4,\Ameas_6$ in the setting basis\\
and record the outcome  $u_{\{2,4,6\}}$};
\draw[signal] (corr.south) -- (measure.north);
\node[measurebox,fill=llmrevision!10] (condition) at (4.7,2.45)
 {Condition on $\Omega(u)$, so the state is now given by $\measuredsource{u}_{A_1^6 (A')_1^6}$, \\
 which has tensor product structure (\cref{lemma:theblockstates,lemma:measuretogettensorproduct})};
\draw[signal] (measure.south) -- node[right,font=\scriptsize] {$\measuredsource{1}$} (condition.north);
\foreach \k/\x in {1/.7,2/4.7,3/8.7}{
 \node[sourcebox,minimum width=2.8cm] (s\k) at (\x,.95)
 {$\measuredsource{u}_{\blockversion{A}_{\k,1}}$};
 \draw[signal] (\x,2.02) -- (s\k.north);
}
\node at (2.7,.95) {$\otimes$};
\node at (6.7,.95) {$\otimes$};

\foreach \q/\x in {1/0,2/1.4,3/4,4/5.4,5/8,6/9.4}{
 \node[roundbox] (n\q) at (\x,-1.2) {$\onerndchann_{\q}$};
 \draw[signal] (\x,.62) -- node[pos=.23,right] {$A_{\q}$} (n\q.north);
}
\foreach \k/\first/\last in {1/1/2,2/3/4,3/5/6}{
 \coordinate (join\k) at ($(n\last.east)+(.35,0)$);
 \draw[thick,rounded corners=4pt]
 ($(n\first.north west)+(-.25,.72)$)
 rectangle ($(join\k)+(.18,-1.2)$);
 \node at ($(n\first.north)+(.8,1.0)$) {$\superchannel{\k}{1}$};
 \draw[signal] (n\first.east) -- (n\last.west);
 \draw[signal] (n\first.south) -- ++(0,-1.4)
 node[below] {$S_{\first}\CP_{\first}$};
 \draw (n\last.east) -- (join\k);
 \draw[signal] (n\last.south)
 -- node[left,font=\scriptsize] {$S_{\last}\CP_{\last}$} ++(0,-.65)
 -| (join\k);
 \node[junction] at (join\k) {};
}
\draw[signal] (condition.west) -- (-2.7,2.45) -- (-2.7,-1.2)
 -- node[above] {$\blockversion{E}_{0,1}=E_0$} (n1.west);
\draw[signal] (join1) -- node[above,font=\scriptsize] {$\blockversion{E}_{1,1}$} (n3.west);
\draw[signal] (join2) -- node[above,font=\scriptsize] {$\blockversion{E}_{2,1}$} (n5.west);
\draw[signal] (join3) -- (12,-1.2) node[right] {$\blockversion{E}_{3,1}$};
\node[align=center,font=\scriptsize] at (5,-3.95)
 {$\Ameas_2,\Ameas_4,\Ameas_6$ have the fixed values $u_2,u_4,u_6$; \\
 $\blockversion{E}_{3,1}=E_6S_{\{2,4,6\}}\CP_{\{2,4,6\}}$.};
\end{scope}
\end{tikzpicture}
\caption{An example of the grouped registers and channels used in the proof of \cref{thm:source-blockwise-meat}, for $n=6$, $\lcsrc=1$, and $r=1$. Each small box is the full physical round channel $\onerndchann_j=\QKDGmap{j}\circ\attack{j}$ from \cref{fig:meat_attack}. Panel (a) describes the original correlated source. Panel (b) first measures $\Ameas_{u_{\otherinterleavedrounds{r}}}$,  records their outcomes, and then groups together the rounds appropriately. }
\label{fig:Meatsourcecorrelations}
\end{figure}

We now have all the pieces needed to prove a version of the MEAT statement that is applicable to correlations. It is useful to refer to \cref{fig:Meatsourcecorrelations} throughout this subsection. Our proof will proceed in three steps.
\begin{enumerate}
    \item 
For each fixed interleaving $r$ and outcome $u_{\otherinterleavedrounds{r}}$, we have a tensor product structure from \cref{lemma:measuretogettensorproduct,lemma:theblockstates}. We can therefore apply MEAT by grouping the physical rounds in each $\sourceblock{k}{r}$, and obtain a tradeoff function on $\CP_{1,r}^n$ that can be used (loosely speaking) to lower bound $\Halpha(S_{1,r}^n | \CP_{1,r}^n \CP_{1,\neq r}^n S_{1, \neq r}^n E_n)_{\rho_{ | \Omega(u_{\otherinterleavedrounds{r}})}}$. We do not want this tradeoff function to depend on $u_{\otherinterleavedrounds{r}}$. We want the same tradeoff function to act as a lower bound regardless of  $u_{\otherinterleavedrounds{r}}$. This can be easily achieved by performing a suitable worst-case analysis over $u_{\otherinterleavedrounds{r}}$. This is done in \cref{thm:source-blockwise-meat}.
\item The $\kappa$ obtained from the  blockwise application of MEAT involves a blockwise optimization. This can be simplified to a single-round optimization in \cref{lem:source-block-relaxation}.
\item Now that we have a tradeoff function that works for all $u_{\otherinterleavedrounds{r}}$, we can get rid of the conditioning on  $\Omega (u_{\otherinterleavedrounds{r}})$  using standard properties of the {\Renyi} entropy. Moreover, the fact that we perform the $\measmap{j}$ on the $\otherinterleavedrounds{r}$ rounds is immaterial, since those measurement do not affect the final output state as shown in \cref{lem:early-setting-measurement}. This is done in \cref{lemma:gettingridofconditioning}. 
\item 
Finally, we can do the above analysis separately for each value of $r$ and combine everything together using \cref{cor:weighted-multiblock-chain-rule}. The final statement is thus obtained in \cref{theorem:meatwithsourcecorrelations}.

\end{enumerate}

\begin{theorem}[Blockwise MEAT after measuring $\Ameas_{\otherinterleavedrounds{r}}$]
\label{thm:source-blockwise-meat}
Consider the scenario in \cref{fig:meat_attack}, with the  protocol maps $\QKDGmap{j}$ and source-replaced state 
satisfying
\cref{condition:sourcemodel,condition:protocolmapsmeasAj} respectively.
Fix an $r\in[\corrperiod]$, and an outcome
$u_{\otherinterleavedrounds{r}}$, and let
$\alpha\in(1,\infty)$. Write
\begin{equation}
 \rho^{u_{\otherinterleavedrounds{r}}}_{S_1^n \CP_1^n E_n} \defvar
 \bigl(\QKDGmap{n}\circ\attack{n}\circ\cdots\circ
 \QKDGmap{1}\circ\attack{1}\bigr)
[\measuredsource{u_{\otherinterleavedrounds{r}}}_{A_1^n(A')_1^n}].
\end{equation}
For each $k\in[m_r]$, let $\jstart=\selectedround{k}{r}$ be the index of the start of the block $\sourceblock{k}{r}$ 
$\jlast=\min\{j+\lcsrc,n\}$ denote the end (with suitable modifications for the first and last blocks). Define 
\begin{equation}
 \superchannel{k}{r}
\defvar \QKDGmap{\jlast}\circ\attack{\jlast}\circ\cdots\circ
 \QKDGmap{j}\circ\attack{j}
 \in\CPTP(\blockversion{A}_{k,r}\blockversion{E}_{k-1,r},
 S_j\CP_j\blockversion{E}_{k,r}),
\end{equation}
where $\blockversion{A}_{k,r}=A_jA_{j+1}^{\jlast}$,
$\blockversion{E}_{k,r}=E_{\jlast} S_{1,\neq r}^{\jlast}\CP_{1,\neq r}^{\jlast}$. For each selected-round history
$\cP_{1,r}^{j-1}$, let 
$f^{(r)}_{|\cP_{1,r}^{j-1}}$ be a tradeoff function on $\CP_j$. Define an optimization over the whole block via
\begin{equation} 
\begin{aligned}
\kappa^{\mathrm{block},u_{\otherinterleavedrounds{r}}}_{\alpha, k,r}
 \defvar& \inf_{\substack{
 \omega\in\dop{=}(\blockversion{A}_{k,r}
 \blockversion{E}_{k-1,r}\widetilde E)\\
 \omega_{\blockversion{A}_{k,r}} =\measuredsource{u_{\otherinterleavedrounds{r}}}_{\blockversion{A}_{k,r}}}}
 \Halpha[f^{(r)}_{|\cP_{1,r}^{j-1}}](S_j\mid\CP_j\blockversion{E}_{k,r}\widetilde E)
 _{\superchannel{k}{r}[\omega]} \\
\kappa^{\mathrm{block}}_{\alpha, k,r} \defvar &  
 \inf_{u_{\otherinterleavedrounds{r}}} \kappa^{\mathrm{block},u_{\otherinterleavedrounds{r}}}_{\alpha, k,r}
 \label{eq:source-block-kappa-separated}
 \end{aligned}
\end{equation}
where $\widetilde E$ is a purifying register. 
Then the function
\begin{equation}
 \fhatfull^{(r)}(\cP_{1,r}^n)
 \defvar\sum_{k=1}^{m_r}
 \left(
 f^{(r)}_{|\cP_{1,r}^{\selectedround{k}{r}-1}}
 (\cP_{\selectedround{k}{r}})
 +\kappa^{\mathrm{block}}_{\alpha, k,r}
 \right)
 \label{eq:source-block-tradeoff-separated}
\end{equation}
satisfies
\begin{equation}
 \Halpha[\fhatfull^{(r)}]
(S_{\interleavedrounds{r}}\mid\CP_{\interleavedrounds{r}} S_{\otherinterleavedrounds{r}}\CP_{\otherinterleavedrounds{r}}E_n)
 _{\rho^{u_{\otherinterleavedrounds{r}}}}\geq0.
 \label{eq:source-block-meat-separated}
\end{equation}
\end{theorem}

\begin{proof}
By \cref{lemma:measuretogettensorproduct}, the state $\measuredsource{u_{\otherinterleavedrounds{r}}}_{A_1^n(A')_1^n}$ is tensor product on the $\blockversion{A}_{k,r}$ groups or registers. Thus, we can apply MEAT in the form of \cref{thm:meatqkd}, after grouping together all the rounds in a given block. For a block starting at round $j=\selectedround{k}{r}$, we treat $S_j$ as the secret output register, $\CP_j$ as the classical register on which the round-wise tradeoff function is defined, and treat all other output registers as Eve's side information $\blockversion{E}_{k,r}$. Thus, applying MEAT yields exactly the global tradeoff function defined in the theorem statement; with the only difference being there is no infimum over $u_{\otherinterleavedrounds{r}}$. However, that infimum can be added trivially using the monotonicity of $f$-weighted {\Renyi} entropy wrt $f$ (see \cref{lemma:Monotonicity_f}). This concludes the proof.
\end{proof}

The $\kappa$ obtained in the theorem above is given by an optimization that involves the input registers of the entire block,
which makes it cumbersome to evaluate numerically, especially for large values of $\lcsrc$. It also complicates a further step that involves taking an additional infimum over all attack channels, which is necessary for QKD security analysis. To address this, we prove the following lemma, which lower bounds it via a single-round optimization.

\begin{lemma}[Relaxing the blockwise optimization to one round]
\label{lem:source-block-relaxation}
In the setting of \cref{thm:source-blockwise-meat},
fix $k$, $r$, $j = \selectedround{k}{r}$, and any tradeoff function $f^{(r)}_{|\cP_{1,r}^{j-1}}$ on $\CP_j$.
Then
\begin{equation}
 \inf_{u_{\otherinterleavedrounds{r}}} \quad \kappa^{\mathrm{block},u_{\otherinterleavedrounds{r}}}_{\alpha,k,r}
 \geq \inf_{u_{\otherinterleavedrounds{r}}} \quad 
 \kappafuncgeneric{f^{(r)}_{|\cP_{1,r}^{j-1}}}{\measuredsource{u_{\otherinterleavedrounds{r}}}_{A_j}}
 {\QKDGmap{j}}{\attack{j}},
 \label{eq:source-block-relaxation-separated}
\end{equation}
where the right-hand side is defined in 
\cref{thm:meatqkd}.
\end{lemma}
\begin{proof}
Fix a $u_{\otherinterleavedrounds{r}}$, and write $f=f^{(r)}_{|\cP_{1,r}^{j-1}}$. It is then enough to show that 
\begin{equation} \label{eq:blockrelaxationtempone}
    \inf_{\substack{
 \omega\in\dop{=}(\blockversion{A}_{k,r}
 \blockversion{E}_{k-1,r}\widetilde E)\\
 \omega_{\blockversion{A}_{k,r}} =\measuredsource{u_{\otherinterleavedrounds{r}}}_{\blockversion{A}_{k,r}}}}
 \Halpha[f](S_j\mid\CP_j\blockversion{E}_{k,r}\widetilde E)
 _{\superchannel{k}{r}[\omega]} \geq  \inf_{\substack{
 \omega\in\dop{=}(A_j E_{j-1} \widehat{E})\\
 \omega_{A_j} =\measuredsource{u_{\otherinterleavedrounds{r}}}_{A_j}}}
 \Halpha[f](S_j\mid\CP_jE_j \widehat{E})
 _{\QKDGmap{j} \circ \attack{j} [\omega]},
\end{equation}
where $\widetilde{E},\widehat{E}$ are purifying registers, since the required statement (\cref{eq:source-block-relaxation-separated}) follows from the above inequality by simply taking an infimum over $u_{\otherinterleavedrounds{r}}$ on both sides. To show \cref{eq:blockrelaxationtempone}, consider a state $w^*_{\blockversion{A}_{k,r} \blockversion{E}_{k-1,r} \widetilde{E}}$  in the feasible set of the optimization on the LHS. It is sufficient to show that there exists a feasible state on the RHS that yields a lower value of the objective function. We will show this as follows.

Let us rewrite $w^*_{\blockversion{A}_{k,r} \blockversion{E}_{k-1,r} \widetilde{E}}$ as $w^*_{A_j E_{j-1} R_\mathrm{extra} \widetilde{E}}$, where $R_\mathrm{extra} = S_{1,\neq r}^{j-1} \CP_{1,\neq r}^{j-1} A_{j+1}^{\jlast}$, where $\jlast = \min\{j+\lcsrc,n\}$ is the index of the last physical round in that block. After the maps from round $j$ are applied, this state becomes
\begin{equation}
    \omega^*_{S_j \CP_j R_\mathrm{extra} E_j } = \QKDGmap{j} \circ \attack{j} \left[ w^*_{A_j E_{j-1} R_\mathrm{extra} \widetilde{E}} \right]
\end{equation}
and the final state $w^*_{S_j \CP_j \blockversion{E}_{k,r} \widetilde{E}}$ is obtained via some additional CPTP map $\mathcal{E}_\mathrm{extra} \in \CPTP(R_{\mathrm{extra}} E_j,\blockversion{E}_{k,r})$. (This map acts trivially on some of the subsystems of $R_{\mathrm{extra}}$, but we do not need to utilize that fact in our proof.)
Thus, using data processing (\cref{lemma:DPIfweighted}), we have
\begin{equation}
   \Halpha[f](S_j\mid\CP_j\blockversion{E}_{k,r}\widetilde E)
 _{\superchannel{k}{r}[\omega^*]} =   \Halpha[f](S_j\mid\CP_j\blockversion{E}_{k,r}\widetilde E)
 _{\mathcal{E}_\mathrm{extra} \circ \QKDGmap{j} \circ \attack{j} [\omega^*]} \geq \Halpha[f](S_j\mid\CP_j E_j R_\mathrm{extra} \widetilde{E} )
 _{ \QKDGmap{j} \circ \attack{j} [\omega^*]} 
\end{equation}
Let $w'_{A_j E_{j-1} \widehat{E}}$ be the same state as $w^*_{A_j E_{j-1} R_\mathrm{extra} \widetilde{E}}$, where we identify $\widehat{E} = R_\mathrm{extra} \widetilde{E} $. By construction, $w'_{A_j} = \Tr_{A_{j+1}^{\jlast}} \left[ \omega^*_{\blockversion{A}_{k,r}} \right]$, and thus belongs to the feasible state on the RHS of \cref{eq:blockrelaxationtempone}. This proves the required result.

\end{proof}

In the next lemma we remove the conditioning on $\Omega(u_{\otherinterleavedrounds{r}})$ from the statements we have obtained.

\begin{lemma} \label{lemma:gettingridofconditioning}
Consider the scenario in \cref{fig:meat_attack} and \cref{thm:source-blockwise-meat,lemma:measuretogettensorproduct}, where the correlated source satisfies \cref{condition:sourcemodel} and the protocol maps satisfy \cref{condition:protocolmapsmeasAj}. Fix an $r\in[\corrperiod]$, let
$\alpha\in(1,\infty)$, and let $\fhatfull^{(r)}$ be a tradeoff function on $\CP_{\interleavedrounds{r}}$ such that
\begin{equation} \label{eq:combiningstart}
     \Halpha[\fhatfull^{(r)}]
(S_{\interleavedrounds{r}}\mid\CP_{\interleavedrounds{r}} S_{\otherinterleavedrounds{r}}\CP_{\otherinterleavedrounds{r}}E_n)
 _{\QKDGmap{n} \circ \attack{n} \dots \circ \QKDGmap{1} \circ \attack{1} \left[ \measuredsource{u_{\otherinterleavedrounds{r}}}_{A_1^n (A')_1^n} \right]}\geq0.
\end{equation}
for all $u_{\otherinterleavedrounds{r}}$. Then the same tradeoff function satisfies
\begin{equation} \label{eq:combiningstartsecond}
    \Halpha[ \fhatfull^{(r)}](S_{\interleavedrounds{r}}\mid\CP_{\interleavedrounds{r}} S_{\otherinterleavedrounds{r}}\CP_{\otherinterleavedrounds{r}}E_n)_{ \QKDGmap{n} \circ \attack{n} \circ \dots \QKDGmap{1} \circ \attack{1} \left[ \sigma^\mathrm{corr}_{A_1^n (A')_1^n} \right]} \geq 0,
\end{equation}
\end{lemma}
\begin{proof}
For a given $r$, consider a scenario where we implement the measurements $\measmap{\otherinterleavedrounds{r}}$. Let the outcomes be $u_{\otherinterleavedrounds{r}}$ and consider the input state to the sequence of channels to be $\sum_{u_{\otherinterleavedrounds{r}}} \Pr(\Omega(u_{\otherinterleavedrounds{r}})) \ketbra{u_{\otherinterleavedrounds{r}}}_U \otimes  \measuredsource{u_{\otherinterleavedrounds{r}}}_{A_1^n (A')_1^n}$ (\cref{eq:measuredstateddefined}) instead of the $\sigma^\mathrm{corr}$. Let $\nu_{S_{\interleavedrounds{r}} \CP_{\interleavedrounds{r}}
 S_{\otherinterleavedrounds{r}}\CP_{\otherinterleavedrounds{r}}E_n U } \defvar \QKDGmap{n} \circ \attack{n} \dots \QKDGmap{1}\circ \attack{1} \left[ \sum_{u_{\otherinterleavedrounds{r}}} \Pr(\Omega(u_{\otherinterleavedrounds{r}})) \ketbra{u_{\otherinterleavedrounds{r}}}_U \otimes  \measuredsource{u_{\otherinterleavedrounds{r}}}_{A_1^n (A')_1^n} \right] $.

Now, \cref{eq:combiningstart} directly gives us (from \cref{def:fweighted}) that
\begin{equation}
    \sum_{\cP_{1,r}^n} \Pr( \Omega( \cP_{1,r}^n ) | \Omega(u_{\otherinterleavedrounds{r}}) ) 2^ { \left( \frac{1-\alpha}{\alpha} \right) \left(  \Halpha(S_{\interleavedrounds{r}}\mid
 S_{\otherinterleavedrounds{r}}\CP_{\otherinterleavedrounds{r}}E_n )_{\nu | \Omega(\cP_{1,r}^n) \wedge \Omega(u_{\otherinterleavedrounds{r}})} - \fhatfull^{(r)} \right) }  \leq 1
\end{equation}

We can multiply with $\Pr(\Omega(u_{\otherinterleavedrounds{r}}))$ and sum over all $u_{\otherinterleavedrounds{r}}$ to obtain

\begin{align}
 1 & \geq \sum_{ u_{\otherinterleavedrounds{r}}}  \Pr( \Omega( u_{\otherinterleavedrounds{r}}))\sum_{\cP_{1,r}^n} \Pr( \Omega( \cP_{1,r}^n ) | \Omega(u_{\otherinterleavedrounds{r}}) )2^{ \left(  \frac{1-\alpha}{\alpha} \right) \left(  \Halpha(S_{\interleavedrounds{r}}\mid
S_{\otherinterleavedrounds{r}}\CP_{\otherinterleavedrounds{r}}E_n )_{\nu | \Omega(\cP_{1,r}^n) \wedge \Omega(u_{\otherinterleavedrounds{}})} - \fhatfull^{(r)} \right) }  \label{eq:combiningtemp1} \\
 & = \sum_{\cP_{1,r}^n} \Pr( \Omega( \cP_{1,r}^n ) ) \sum_{ u_{\otherinterleavedrounds{r}}}  \Pr( \Omega( u_{\otherinterleavedrounds{r}}) | \Omega( \cP_{1,r}^n ) )  2^{ \left(  \frac{1-\alpha}{\alpha} \right) \left(  \Halpha(S_{\interleavedrounds{r}}\mid
S_{\otherinterleavedrounds{r}}\CP_{\otherinterleavedrounds{r}}E_n )_{\nu | \Omega(\cP_{1,r}^n) \wedge \Omega(u_{\otherinterleavedrounds{}})} - \fhatfull^{(r)} \right) }  \label{eq:combiningtemp2} \\
& = \sum_{\cP_{1,r}^n} \Pr( \Omega( \cP_{1,r}^n ) )     2^{ \left(  \frac{1-\alpha}{\alpha} \right) \left(  \Halpha(S_{\interleavedrounds{r}}\mid
S_{\otherinterleavedrounds{r}}\CP_{\otherinterleavedrounds{r}}E_n U )_{\nu | \Omega(\cP_{1,r}^n) )} - \fhatfull^{(r)} \right) }  \label{eq:combiningtemp3} \\
 &\geq  \sum_{\cP_{1,r}^n} \Pr( \Omega( \cP_{1,r}^n ) )2^{ \left(  \frac{1-\alpha}{\alpha} \right) \left(  \Halpha(S_{\interleavedrounds{r}}\mid 
S_{\otherinterleavedrounds{r}}\CP_{\otherinterleavedrounds{r}}E_n )_{\nu | \Omega(\cP_{1,r}^n) )} - \fhatfull^{(r)} \right) } \label{eq:combiningtemp4} \\
&=\label{eq:combiningtemp5}   2^{  \left( \frac{1-\alpha}{\alpha}  \right)\left(  \Halpha[\fhatfull^{(r)}]
(S_{\interleavedrounds{r}}\mid\CP_{\interleavedrounds{r}}
S_{\otherinterleavedrounds{r}}\CP_{\otherinterleavedrounds{r}}E_n)_{ \nu }  \right)} 
 \end{align}
where we used Bayes's rule for \cref{eq:combiningtemp2},\cite[Proposition 5.1]{tomamichel_quantum_2016} for \cref{eq:combiningtemp3}, data processing (\cref{lemma:DPIfweighted} for \cref{eq:combiningtemp4} and \cref{def:fweighted} for \cref{eq:combiningtemp5}.

The proof now follows from the fact that 
\begin{align}
    \nu_{S_{\interleavedrounds{r}} \CP_{\interleavedrounds{r}}
 S_{\otherinterleavedrounds{r}}\CP_{\otherinterleavedrounds{r}}E_n  } &= \QKDGmap{n} \circ \attack{n} \dots \QKDGmap{1}\circ \attack{1} \left[ \sum_{u_{\otherinterleavedrounds{r}}} \Pr(\Omega(u_{\otherinterleavedrounds{r}})) \measuredsource{u_{\otherinterleavedrounds{r}}}_{A_1^n (A')_1^n} \right] \\
 &=    \QKDGmap{n} \circ \attack{n} \dots \QKDGmap{1}\circ \attack{1} \circ \measmap{\otherinterleavedrounds{r} }\left[\sigma^\mathrm{corr}_{A_1^n (A')_1^n}  \right] \\
 & = \QKDGmap{n} \circ \attack{n} \dots \QKDGmap{1}\circ \attack{1} \circ \left[\sigma^\mathrm{corr}_{A_1^n (A')_1^n}  \right]
\end{align}
where the first equality follows from the definition, the second equality follows from \cref{eq:measuredstateddefined}, and the final inequality follows from the fact that the $\measmap{}$ does not affect the final output state (\cref{condition:protocolmapsmeasAj}). This concludes the proof. 
\end{proof}

We can combine the result of the interleaved protocols together into the following theorem which can be directly applied for analyzing QKD with correlated sources.

\begin{theorem}[MEAT for QKD with correlated sources]
\label{theorem:meatwithsourcecorrelations}
Consider the scenario in \cref{fig:meat_attack}, and suppose that the correlated source satisfied \cref{condition:sourcemodel} and the protocol maps satisfy \cref{condition:protocolmapsmeasAj}. Fix $r\in[L]$, and for each $k\in[m_r]$, define a tradeoff function $f^{(r)}_{|\cP_{1,r}^{\selectedround{k}{r}-1}}$ on $\CP_j$.  Let $\alphafinal,\beta_1,\ldots,\beta_L>1$ satisfy
\begin{equation} \label{eq:alphafinaldefined}
\frac{\alphafinal}{\alphafinal-1}
=\sum_{r=1}^{L}\frac{\beta_r}{\beta_r-1},
\end{equation}
Define global normalized tradeoff functions on $\CP_{1,r}^n$ and $\CP_{1}^{n}$ respectively via
\begin{equation} 
\begin{aligned}
 \fhatfull^{(r)}(\cP_{1,r}^n)
 &\defvar\sum_{k=1}^{m_r}\left(
f^{(r)}_{|\cP_{1,r}^{\selectedround{k}{r}-1}} (\cP_{\selectedround{k}{r}})
 +\inf_{ \hat{u}_{\otherinterleavedrounds{r}}}
\kappafuncgeneric{f^{(r)}_{|\cP_{1,r}^{\selectedround{k}{r}-1}}}{\measuredsource{\hat{u}_{\otherinterleavedrounds{r}}}_{A_{\selectedround{k}{r}}}}
 {\QKDGmap{\selectedround{k}{r}}}{\attack{\selectedround{k}{r}}}[\beta_r]
 \right), \\
  \fhatfull(\cP_{1}^n) &\defvar \sum_{r=1}^{L}  \fhatfull^{(r)}(\cP_{1,r}^n)
 \label{eq:fhatfulldefinedsource}
\end{aligned}
\end{equation}
where $\kappa$ is the single-round quantity in \cref{thm:meatqkd}. 
Then, 
\begin{equation}
    \Halpha[ \fhatfull]<\alphafinal>(S_{1}^n | \CP_{1}^n E_n)_{ \QKDGmap{n} \circ \attack{n} \circ \dots \QKDGmap{1} \circ \attack{1} \left[ \sigma^\mathrm{corr}_{A_1^n (A')_1^n} \right]} \geq 0.
\end{equation}
Moreover, the above inequality remains true even if the single-round $\kappa$ optimization in \cref{eq:fhatfulldefinedsource} is replaced by the blockwise version from \cref{eq:source-block-kappa-separated}.
\end{theorem}

\begin{proof}
The proof follows straightforwardly by applying \cref{lemma:gettingridofconditioning,thm:source-blockwise-meat} and combining the resulting bounds using \cref{cor:weighted-multiblock-chain-rule}. Doing so results in the $\kappa$ being given by the blockwise version from \cref{eq:source-block-kappa-separated}. This can be replaced with the single-round version from \cref{eq:fhatfulldefinedsource} using \cref{lem:source-block-relaxation}.
    
\end{proof}
We now discuss several aspects of applying these results to
correlated source imperfections.
\subsection{Discussion and Examples} \label{subsec:discussionsource}

Notice that our proof naturally yields the blockwise $\kappa$ term
entering \cref{eq:source-block-kappa-separated}, which we subsequently
lower-bound by a single-round optimization to obtain
\cref{eq:fhatfulldefinedsource}. In this relaxation, we effectively
give Eve the $A_q$ systems of all rounds in the block after the
selected round $\selectedround{k}{r}$.
The setting registers $\Ameas_q$ are already fixed to pure states
by conditioning on $u_{\otherinterleavedrounds{r}}$, so including
these registers provides no additional information to Eve. However, the shield registers $\Ashield_q$
remain inside Alice's laboratory in the actual protocol, whereas
the relaxation includes them in Eve's side-information.
Thus, although the current round's shield $\Ashield_{\selectedround{k}{r}}$ remains
a part of Alice's input to the protocol map (and is not given to Eve),  we give Eve the shield systems of
the later rounds in the same block. Thus, this step is not tight, and a more refined analysis may be able to tighten this part of our analysis. 
We therefore retain both the blockwise and single-round
formulations.\footnote{One can imagine interpolating between the blockwise formulation and the single-round formulation, which would result in a tradeoff of tightness over numerical cost. However we do not explore this idea further here. }

Direct evaluation of the blockwise optimization may only be
practical for very small values of $\lcsrc$, so we expect the
single-round formulation to be more useful numerically. (Furthermore, one would also additionally need to optimize over all possible attacks by Eve.)
The loss in performance due to this relaxation may be small when the future shield systems convey little additional information about the selected
round, but this requires a quantitative assessment for the
source model under consideration, which we leave to future work Ref.~\cite{inprep_correlatedjerome}.

Second, the single-round bound involves an infimum over
$u_{\otherinterleavedrounds{r}}$. By \cref{lemma:theblockstates},
the marginal $\measuredsource{u_{\otherinterleavedrounds{r}}}_{A_j}$
depends only on the setting choices in nearby rounds.
Consequently, it suffices to optimize over those entries of
$u_{\otherinterleavedrounds{r}}$, rather than the entire string.
The resulting problem is a single-round $\kappa$ optimization
with an additional optimization over a set of possible 
marginals constraints on the $A_j$ system. This has the same form as the problem arising from
independent source imperfections when the source-replaced
marginal is a tensor product but its individual factors are not
known exactly. Thus, we have reduced the
analysis of correlated source imperfections to independent imperfections. The methods of
Ref.~\cite{wiesemann_securityframework_2026}, which studies a variety of practical independent source imperfections, can therefore be
applied once the resulting set of admissible marginals has been
characterized in the required form.
Note that Ref.~\cite{curraslorenzo_phaseerrorcorrelated_2026} obtains a similar
reduction for security analyses based on phase error
estimation.

Third, notice that combining the interleaved subsequences changes the
{\Renyi} parameter. Setting $\beta_r=\alpha$ for every $r$
in \cref{eq:alphafinaldefined} gives
\begin{equation}
    \frac{\alphafinal}{\alphafinal-1}
    = L\frac{\alpha}{\alpha-1},
    \qquad
    \alphafinal
    = \frac{L\alpha}{1+(L-1)\alpha}.
\end{equation}
Consequently, the privacy-amplification penalty in the
key length expression (\cref{eq:keylengthformula}) \cite{BDR3,kamin_renyi_2025} becomes
\begin{equation}
    \frac{\alphafinal}{\alphafinal-1}
    \log\frac{1}{\varepsilon_{\mathrm{PA}}}
    =
    L\frac{\alpha}{\alpha-1}
    \log\frac{1}{\varepsilon_{\mathrm{PA}}}.
\end{equation}
For a fixed choice of $\alpha$, this term is therefore
multiplied by $L$. In particular, if
$\alpha=1+c/\sqrt{n}$ for a constant $c>0$, then
\begin{equation}
    \frac{\alphafinal}{\alphafinal-1}
    \log\frac{1}{\varepsilon_{\mathrm{PA}}}
    =
    L\left(\frac{\sqrt{n}}{c}+1\right)
    \log\frac{1}{\varepsilon_{\mathrm{PA}}}.
\end{equation}
For fixed $L$, $c$, and $\varepsilon_{\mathrm{PA}}$, this results in a factor of $L$ worsening of the $O(\sqrt{n})$ penalty to the key length. (Note that the reduction in the value of $\fhatfull$ due to the additional infimum over all the marginals is a separate penalty due to correlations.)

Fourth, our source model directly accommodates mixed emitted
states, whereas the source model and reduction in
Ref.~\cite{curraslorenzo_phaseerrorcorrelated_2026} are formulated
for pure emitted states. Thus, unlike Ref.~\cite{curraslorenzo_phaseerrorcorrelated_2026} our results can be directly applied to intensity correlations for phase randomized laser pulses. 
 As discussed in
\cref{sec:combined}, the source correlation analysis can be
combined with the detector correlation results, allowing both
effects to be treated within the same security analysis under
the stated assumptions.

Finally, one may wish to analyze sources with unbounded correlation
length. A possible approach is to introduce a finite cutoff
$\lcsrc$ and construct an approximate source model in which
correlations beyond this cutoff are neglected. Our results can
then be applied to the truncated model. To obtain a security
guarantee for the actual source, one must additionally bound the
trace distance between the actual and truncated source-replaced
states over the full protocol, as is done in \cite[Lemma 1]{curraslorenzo_phaseerrorcorrelated_2026}. 
The usefulness of this approach depends on obtaining 
sufficiently small truncation error from the source model, which is typically possible as long as the correlations decay sufficiently fast.

\section{Combining Source and Detector Correlations}
\label{sec:combined}

We now combine the arguments of
\cref{theorem:meatwithsourcecorrelations,thm:meat-effective-history}
to obtain a statement applicable to QKD with source correlations,
detector correlations, and iterative sifting.

\begin{theorem}[MEAT for source and detector correlations and iterative sifting]
\label{theorem:combinedstatement}
Consider the sequence of channels in
\cref{fig:meat_attack_memory}, with attack channels
$
    \attack{j}\in\CPTP(E_{j-1},B_jE'_j)
$
and protocol maps
$
    \QKDGmap{j}
    \in\CPTP(A_jB_j\Mem_{j-1},
             S_j\CP_j\CPhat_j\Mem_j),
$
where $\Mem_0$ be initialized in a fixed classical state. Thus, the sequence of channels is given by
\begin{equation}
    \rho_{S_1^n\CP_1^nE_n\Mem_n}
    =
    \left(\QKDGmap{n} \circ \attack{n} \dots \circ \QKDGmap{1} \circ \attack{1} \right)
    \left[
        \sourcesymbol^{\mathrm{corr}}_{A_1^nE_0}
        \otimes\ketbra{m_0}_{\Mem_0}
    \right].
\end{equation}
Supposed the source-replaced state $\sourcesymbol^{\mathrm{corr}}_{A_1^nE_0}$ satisfies
\cref{condition:sourcemodel}, and 
that the protocol maps satisfy
\cref{condition:protocolmapsmeasAj,condition:classicalmemoryandmap,condition:creatingnextmemory,condition:coarsegrainingmemory}. In particular,
\begin{equation}
    \begin{aligned}   
    \QKDGmap{j}
    =&
    \QKDGmap{j}\circ
    (\measmap{j}\otimes\id_{B_j\Mem_{j-1}}), \\
    \QKDGmap{j}
    =&
    \createhistory{j}\circ
    \effectiveQKDGmap{j}\circ
    \coarsegrainedhistory{j}, \\
     \effectiveQKDGmap{j}
    \in&\CPTP(A_jB_j\EffMem_j,S_j\CP_j\CPhat_j),
    \end{aligned}
\end{equation}
where $\EffMem_j$ stores the coarse-grained memory, and the various maps are defined in the stated conditions.  Let $\alphafinal,\beta_1,\ldots,\beta_L>1$ satisfy
\begin{equation}
    \frac{\alphafinal}{\alphafinal-1}
    =
    \sum_{r=1}^{L}\frac{\beta_r}{\beta_r-1}.
    \label{eq:combined-renyi-parameters}
\end{equation}
For each $r\in[L]$, $k\in[m_r]$, and selected-round history
$\cP_{1,r}^{\selectedround{k}{r}-1}$, let
$f^{(r)}_{|\cP_{1,r}^{\selectedround{k}{r}-1}}$
be a tradeoff function on $\CP_{\selectedround{k}{r}}$.
Define tradeoff functions on $\CP_{1,r}^n$ and $\CP_1^n$ by
\begin{equation}
\begin{aligned}
    \fhatfull^{(r)}(\cP_{1,r}^n)
    &\defvar
    \sum_{k=1}^{m_r}\Bigl(
    f^{(r)}_{|\cP_{1,r}^{\selectedround{k}{r}-1}}
       (\cP_{\selectedround{k}{r}}) + 
    \inf_{\hat u_{\otherinterleavedrounds{r}}}
    \kappafuncgeneric
        {f^{(r)}_{|\cP_{1,r}^{\selectedround{k}{r}-1}}}
        {\measuredsource{\hat u_{\otherinterleavedrounds{r}}}
            _{A_{\selectedround{k}{r}}}}
        {\effectiveQKDGmap{\selectedround{k}{r}}}
        {\attack{\selectedround{k}{r}}}[\beta_r]
    \Bigr),
    \\
    \fhatfull(\cP_1^n)
    &\defvar
    \sum_{r=1}^{L}\fhatfull^{(r)}(\cP_{1,r}^n)
\end{aligned}
\label{eq:fhatfulldefinedsourcedetcombined}
\end{equation}
where $\measuredsource{\hat u_{\otherinterleavedrounds{r}}}
            _{A_{\selectedround{k}{r}}}$ is defined in
\cref{lemma:theblockstates}, and the effective single-round
optimization is defined as in \cref{eq:effectivekappadefined}.
Then
\begin{equation}
    \Halpha[\fhatfull]<\alphafinal>
        (S_1^n\mid\CP_1^nE_n)_\rho
    \geq
    \Halpha[\fhatfull]<\alphafinal>
        (S_1^n\mid\CP_1^nE_n\Mem_n)_\rho
    \geq 0.
\end{equation}
\end{theorem}

\begin{proof}
Note that due to data-processing (\cref{lemma:DPIfweighted}), it suffices to show that $    \Halpha[\fhatfull]<\alphafinal>
        (S_1^n\mid\CP_1^nE_n\Mem_n)_\rho\geq 0$. The proof follows by combining the proofs of \cref{theorem:meatwithsourcecorrelations,lem:effective-history-reduction}. 
We first repeat the proof of
\cref{theorem:meatwithsourcecorrelations}, including $\Mem_j$
in the side-information register after round $j$. With this identification, we can repeat the proof of \cref{theorem:meatwithsourcecorrelations} and obtain a global tradeoff function that is similar to the one required in the theorem statement, except that the $\kappa$ term is now given by 
\begin{equation}
    \kappafuncgeneric
        {f^{(r)}_{|\cP_{1,r}^{j-1}}}
        {\measuredsource{\hat u_{\otherinterleavedrounds{r}}}_{A_j}}
        {\QKDGmap{j}}{\attack{j}}[\beta_r],
    \qquad j=\selectedround{k}{r}.
\end{equation}
This quantity is defined as in \cref{eq:historykappadefined}.

Now, the proof of \cref{lem:effective-history-reduction} can be repeated without any modification, to obtain 
\begin{equation}
    \kappafuncgeneric
        {f^{(r)}_{|\cP_{1,r}^{j-1}}}
        {\measuredsource{\hat u_{\otherinterleavedrounds{r}}}_{A_j}}
        {\effectiveQKDGmap{j}}{\attack{j}}[\beta_r]
    \leq
    \kappafuncgeneric
        {f^{(r)}_{|\cP_{1,r}^{j-1}}}
        {\measuredsource{\hat u_{\otherinterleavedrounds{r}}}_{A_j}}
        {\QKDGmap{j}}{\attack{j}}[\beta_r].
\end{equation}
Using the monotonicity of the $f$-weighted {\Renyi} entropy wrt $f$ (\cref{lemma:Monotonicity_f}), we obtain
\begin{equation}
    \Halpha[\fhatfull]<\alphafinal>
    (S_1^n\mid\CP_1^nE_n\Mem_n)_\rho
    \geq 0.
\end{equation}
for the $\fhatfull$ defined in the theorem statement. This concludes the proof. 
\end{proof}

\section{Plots} \label{sec:plots}
In this section we will apply our results to the qubit BB84 protocol with source and detector correlations. Our focus is on highlighting that our methods produce competitive key rates. Thus, we do not do a full security proof, and instead rely on the analysis already done in prior works \cite{BDR3} and extended to independent imperfections  \cite{wiesemann_securityframework_2026}. We first state the correlated source model, then the detector model, and then the additional details of the protocol we analyze.

\paragraph*{Source Model:} Alice uses independent randomness to select one out of the four possible qubit BB84 states to be sent in a given round. We assume that the only imperfection present is that the polarization of the single-photon is correlated with previous choices. Thus, in each round, Alice prepares
\begin{equation}
    \begin{aligned}
        \ket{\Psi_{i_j}^{(j,i_{j-\lcsrc}^{j-1})}}_{A'_j} &=  \cos (\theta_{i_j}^{(j,i_{j-\lcsrc}^{j-1})})  \ket{0} + \sin(\theta_{i_j}^{(j,i_{j-\lcsrc}^{j-1})}) \ket{1} \\
        \theta_{i_j}^{(j,i_{j-\lcsrc}^{j-1})} &= \hat{\theta}_{i_j}  + \sum_{l=1}^{\min\{\lcsrc,j-1\}} \delta_l (i_{j-l}), \qquad \qquad (\hat\theta_{0},\hat\theta_{1},\hat\theta_{2},\hat\theta_{3}) =  (0,\pi/2,\pi/4,-\pi/4).
    \end{aligned}
\end{equation}
where $\hat{\theta}$ denotes the values for an uncorrelated source, and $\theta_{i_j}^{(j,i_{j-\lcsrc}^{j-1})}$ denotes a perturbed version of the uncorrelated value, where the perturbation is controlled by $\delta_l$ and depends on previous choices. We assume that the correlations behave as
\begin{equation}
 |\delta_l(x)-\delta_l(x')|\leq\sqrt{\xi_l},\qquad
 \xi_l=\xi_1 e^{-C_\mathrm{decay}(l-1)}\quad(1\leq l\leq\lcsrc),\qquad
 \delta_l(x)=0\quad(l>\lcsrc),
 \label{eq:cbb84-decay}
\end{equation}
where $\xi_l$ denotes the strength of the correlations at length $l$, and this decays exponentially based on the value of $C_\mathrm{decay}$. This is the same model as in  Ref.~\cite{curraslorenzo_phaseerrorcorrelated_2026}, and is supported by experimental results \cite{agulleiro2025modelingcharacterizationarbitraryorder}.

\paragraph*{Detectors:} We assume the canonical model of detectors, an active basis choice detection setup, and assume that correlated effects last for some finite length $\lcdet$, during which the detectors are shut off using appropriate hardware modifications. In this period, Bob announces a special symbol $\discard$ and Alice stores $\bot$ in the $S_j$ register. We assume that when the detector is switched on again, it has unit efficiency and zero dark counts, allowing us to utilize the simple qubit squasher \cite{gittsovich_squashing_2014} (we assume that double-clicks are randomly mapped to single-clicks). Note that the extension to the flag-state squasher \cite{zhang_security_2021} and imperfections such as efficiency mismatch and dark count is immediate, using Ref.~\cite{nahar2025imperfect,wiesemann_securityframework_2026}.

\paragraph*{Other protocol details: } We assume a protocol where Alice and Bob both choose the $\bf{X}$ basis with probability $p_\mathrm{test}$, and all rounds where Alice chooses this basis are used for testing. All other rounds are used for key generation. We assume a one-way error-correction protocol with efficiency $f_\mathrm{EC}=1.16$; thus, the protocol uses $\lambda_\mathrm{EC} (\cP_1^n) = n f_\mathrm{EC} H(S | Y \CP)$ bits upon observing $\cP_1^n$, where the $ H(S | YC)$ term is evaluated on the distribution implied by the announcements in the test rounds, and $Y$ denotes Bob's measurement outcome.\footnote{Recall that in the test rounds, Alice and Bob both announce their entire classical information from that round, which fixes the distribution on $S,Y,\CP$.} We assume error-verification and privacy amplification as described in Ref.~\cite{BDR3}.  We pick the same tradeoff function $f$ in each round. Since we choose the same tradeoff function $f$ in each round, and thus only need to perform a single computation of $\kappa^\mathrm{QKD}$, where the subscript is used to denote that we have performed an additional optimization over all possible attacks by Eve. For more details, we refer the reader to Ref.~\cite{BDR3}. Then, using \cite[Theorem 8.3]{BDR3}, the protocol is $(\varepsilon_\mathrm{PA} + \varepsilon_\mathrm{EV})$-secure if the key length $\ell(\cP_1^n)$ is chosen as
\begin{equation} \label{eq:keylengthformula}
 \ell(\cP_1^n) = \max\left\{ 0 , \floor{ n (\mbf{f} + \kappa^\mathrm{QKD}). \mathrm{freq}(\cP_1^n)  - \lambda_\mathrm{EC}(\cP_1^n) - \ceil{ \log(\frac{1}{\varepsilon_\mathrm{EV}})} -  L \frac{\alpha}{\alpha-1} \log(\frac{1}{\varepsilon_\mathrm{PA}}) + 2 } \right\}.
\end{equation}
where we use $\mathbf{f}$ to denote the tradeoff function $f$ viewed as a vector $\mbf{f}\in\mathbb{R}^{|\mathcal{\CP}|}$, $\mathcal{\CP}$ is the alphabet of the $\CP_j$ registers, and $\mathrm{freq}(\cP_1^n)$ is the frequency distribution of $\cP_1^n$. We note that several techniques for evaluation the optimization in $\kappa$ exist \cite{kamin_renyi_2025,wiesemann_securityframework_2026,navarro_finite_2025}. In this work, since we need to additionally optimize over a set of marginal states, we utilize Ref.~\cite{wiesemann_securityframework_2026}.

\paragraph*{Handling the correlated effects}
We use \cref{theorem:combinedstatement} for our security analysis. To evaluate $\kappa^\mathrm{QKD}$, we need to compute the set of marginal states compatible with our model of correlated source, over all possible values of the measured setting choices; i.e, we need to evaluate the set of possible $\measuredsource{u_{\otherinterleavedrounds{r}}}_{A_j}$ in \cref{lem:source-block-relaxation}. As we have noted earlier, this set appears as the set of possible independent imperfections in our analysis.  Thankfully, we can directly utilize the analysis from Ref.~\cite{curraslorenzo_phaseerrorcorrelated_2026} here, which obtains that the independent imperfections in round $j$ simply has the form that the fidelity between the state that is actually sent (in $(A')_{j}^{j+\lcsrc}$), and some fixed ``target" state without correlations, is lower bounded by  $(1-\varepsilon_\mathrm{src})$, where (see \cref{app:keyrates}):
\begin{align}
 \varepsilon_{\mathrm{src}}
 &\defvar \min\!\left\{1,\xi_1\frac{1-e^{-C_\mathrm{decay}\lcsrc}}{1-e^{-C_\mathrm{decay}}}\right\}.
 \label{eq:cbb84-infidelity}
\end{align}
The target states have the same overlap as the BB84 signal states (as we vary the setting choice $i_j$), and thus the resulting set of marginal states is contained in the one where we analyze the qubit BB84 protocol with independent imperfections, where the emitted state is close in fidelity to the ideal BB84 state. This is exactly the scenario analyzed in Ref.~\cite{wiesemann_securityframework_2026}, whose techniques we utilize here for computing key rates.

The main effect of detector correlations on the simulated statistics
is a rescaling within the test and generation rounds separately. We regard the announcements as follows: Alice first announces whether
the round is a test or generation round, and then Bob and Alice make
their remaining announcements. Crucially,  the test or generation label is announced even when the detectors are shut off (to enable the test statistics to provide information on how often the detectors are shut off). Let $\Ftest(a,b)$ and $\Fgen(a,b)$ denote the joint distributions of
the Alice and Bob announcements in the scenario without correlations, in the test and generation rounds respectively. These are identical to the announcements we would get in the correlated protocol, when the detectors are switched on. If $p_{\mathrm{on}}$
is the probability that the detectors are switched on, then the observed statistics for the correlated detector case are given by
\begin{align} \label{eq:modifystats}
    \Fobstest(a,b)
    &=p_{\mathrm{on}}\Ftest(a,b),
    &
    \Fobsgen(a,b)
    &=p_{\mathrm{on}}\Fgen(a,b),
    \qquad b\ne\discard,
    \nonumber\\
    \Fobstest(a,\discard)
    &=(1-p_{\mathrm{on}})\sum_b\Ftest(a,b),
    &
    \Fobsgen(a,\discard)
    &=(1-p_{\mathrm{on}})\sum_b\Fgen(a,b).
\end{align}
The full announcement distribution is therefore
\begin{equation}
    \begin{aligned}
        \Fobs(\mathrm{test},a,b)
        &=p_{\mathrm{test}}\Fobstest(a,b),\\
        \Fobs(\mathrm{gen},a,b)
        &=(1-p_{\mathrm{test}})\Fobsgen(a,b).
    \end{aligned}
\end{equation}
For the stationary channel simulation, in which every click is
followed by $l_{\mathrm c}^{\mathrm{det}}$ shut-off rounds,
$p_{\mathrm{on}}=(1+l_{\mathrm c}^{\mathrm{det}}
p_{\mathrm{click}})^{-1}$, where $p_{\mathrm{click}}$ is the
probability of a click in a round where detectors are switched on.

\begin{figure}
    \includegraphics[width = \linewidth]{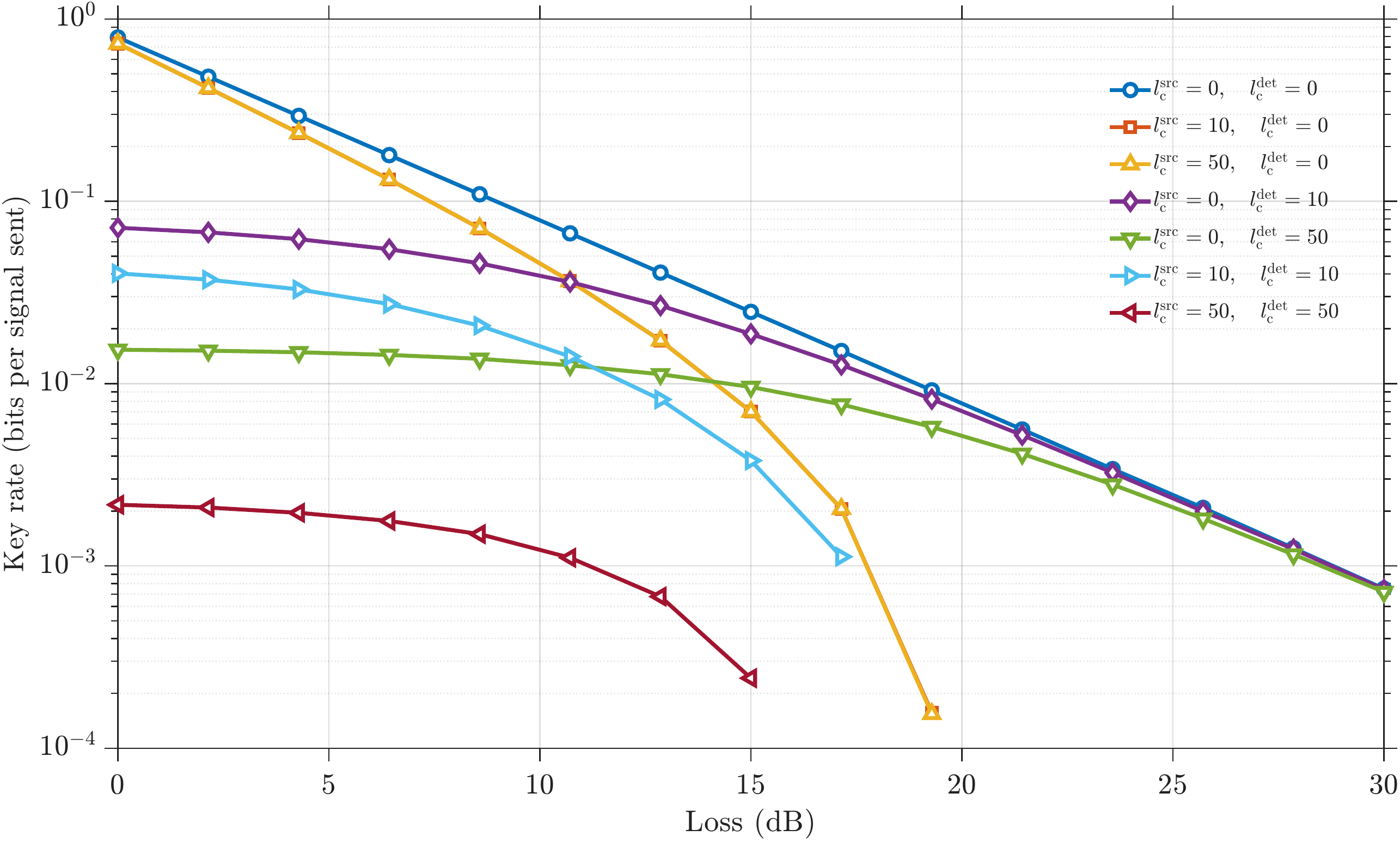}
    \caption{Key rates for the qubit BB84 protocol with correlated sources and detectors. For the results with only source correlations, we find that the effective calculation we perform is similar for many different values of $\lcsrc$ and thus, many values yield similar key rates. For the results with only detector correlations, we find that the penalty is highest in the low-loss regime due to a high volume of detections lost due to the detectors being switched off. 
 } \label{fig:figure}
\end{figure}

With this, we plot key rates\footnote{Note that for the key rate computation we also need to ensure that the $\QKDGmap{j}$ maps appropriately measures the incoming $\EffMem_j$ register and acts accordingly. Furthermore, we also allow Eve to supply the $\EffMem_j$ herself in the optimization, by allowing her to output that register as a part of her attack $\attack{j}$. This only gives her more power. After these replacements, the analysis from Ref.~\cite{BDR3,wiesemann_securityframework_2026} can be applied directly.}   for the qubit BB84 protocol with correlated sources and detectors in \cref{fig:figure}. We set the total number of rounds $n=10^{12}$, $\alpha = 1.001$, and $\varepsilon_\mathrm{EV} = \varepsilon_\mathrm{PA} = 10^{-20} / 2$. We set the probability of testing (which is the same as probability of choosing the $\mathbf{X}$) to be $p_\mathrm{test}=0.1$. For simulating the channel statistics (and only for this task), we make the simplifying assumption that the source is without correlations, and modify the statistics due to detector correlations as described above (\cref{eq:modifystats}). We assume that the channel has loss, and a misalignment of $2^{\circ}$. We set $\xi_1 = 10^{-3}$ and $C_\mathrm{decay} = 12.7$, which are similar to the values plotted in Ref.~\cite{curraslorenzo_phaseerrorcorrelated_2026}.

For the plots with only source correlations, the key rates are
similar over the range of $l_{\mathrm c}^{\mathrm{src}}$ considered.
This is because the correlations decay rapidly, so extending the
correlation length beyond the first few rounds has little effect
on the effective scenario of independent imperfections for which we need to perform our optimization. More formally, we see that the value of $\varepsilon_{\mathrm{src}}$ from \cref{eq:cbb84-infidelity} changes slowly at higher values of $\lcsrc$ (although we stress that we still need to modify the value of the {\Renyi} parameter, as described in \cref{eq:combined-renyi-parameters}). This behavior is
qualitatively similar to that reported in
Ref.~\cite{curraslorenzo_phaseerrorcorrelated_2026}.
For the parameters considered here, the independent imperfection scenario we perform computations for (after using \cref{theorem:combinedstatement}) corresponds to the scenario where the emitted states have fidelity approximately $1-10^{-3}$ with ideal BB84 signal states. The substantially higher key rate for
$l_{\mathrm c}^{\mathrm{src}}=0$ reflects our assumption of an ideal
source, with neither independent nor correlated source imperfections.

For detector correlations, the key-rate penalty is largest in the
low-loss regime and increases with $l_{\mathrm c}^{\mathrm{det}}$.
This is mainly governed by the fact that a lot of  detections are lost in the low-loss regime while the detectors are
shut off. As the channel loss increases, clicks become less frequent, and therefore fewer of them are lost due to the detectors being switched off. This behavior is qualitatively
similar to that observed in Ref.~\cite{wang2025phase}.

\section{Conclusion}
\label{sec:conclusion}

In this work, we derived variants of the MEAT that enable finite-size QKD
security analysis with correlated source imperfections and correlated detector imperfections. The proofs reduce these
settings to applications of the existing MEAT and yield entropy
bounds in terms of single-round optimizations for which numerical
methods are available. The results can be combined, allowing
source and detector correlations to be analyzed together while
retaining on-the-fly announcements and adaptive choices of
tradeoff functions. To the best of our knowledge, this joint
treatment has not been obtained in prior finite-size QKD security
analyses.

For sources, we considered independently chosen settings with
arbitrary dependence of the prepared state on the current setting
and a finite number of preceding settings (although performance is best when the dependence on preceding settings is weak). For detectors, we
assumed that memory effects persist for a finite number of rounds
after a registered click, with all detectors switched off and
those rounds discarded. These settings cover the finite-memory
source and detector models addressed in
Refs.~\cite{curraslorenzo_phaseerrorcorrelated_2026,wang2025phase},
while our analysis also directly accommodates correlated preparation of mixed states. For iterative sifting, we allow any deterministic
stopping rule based on the round index and preceding public
announcements, subject to a fixed maximum number of rounds, for entanglement-based protocols. Extending the source or detector
analysis to unbounded correlation lengths can be obtained by truncating the effect at some finite-length, and then bounding the resulting penalty in terms of trace-distance of the states obtained, as is done in Refs.~\cite{curraslorenzo_phaseerrorcorrelated_2026,nahar_phd_2025}. We applied our results to the qubit BB84 protocol with source and detector correlations, and demonstrate competitive key rates.

Several questions remain open. First, it is important to apply
these results to practical QKD protocols and compute key rates.
Although we discuss their application in
\cref{subsec:discussionsource,subsec:detectorexamples}, and compute key rates in \cref{sec:plots} a complete
security analysis for a practical protocol requires specifying a number of additional details, such as details about the hardware assumptions, and various imperfections. As discussed in the introduction, these details
are beyond the scope of the present work. We leave these
applications to ongoing work~\cite{inprep_correlatedjerome},
which extends the frameworks of
Refs.~\cite{wiesemann_securityframework_2026,BDR3}. 

Second, our detector analysis requires switching off all detectors
and discarding the affected rounds regardless of the strength of
the correlations. When the memory effects are weak and the
measurement POVMs remain close to their usual values, as may
occur for weak afterpulsing, it is natural to ask whether those
rounds could instead contribute to the key. Establishing such a
result is nontrivial: the dependence on previous measurement
outcomes cannot simply be included in Eve's side information
without revealing secret data to Eve, and the secret memory
versions of entropy accumulation impose additional restrictions,
as discussed in \cref{remark:conditioning_imperfections}.
The need to discard rounds independently of the correlation
strength also appears in the phase-error analysis of
Ref.~\cite{wang2025phase}. Developing analyses that account for
the strength of detector correlations, rather than discarding
all affected rounds, is therefore an open problem across these
approaches and could improve practical key rates.

Third, our results rely on specific models of source and detector
correlations. It would be useful to identify weaker assumptions
under which more general correlations can be handled, including
those arising from quantum memory within the devices, such as Ref.~\cite{marwah2024proving}. Such
extensions would broaden the information-theoretic framework,
but their practical value will depend on whether they capture
device behaviour that is relevant to QKD implementations and
not adequately described by the models considered here. For a concrete example, we note while intensity and polarization correlations are
covered by the model considered in this work, phase correlations in phase-randomized sources are not. This is because the phase is not a setting choice when using phase-randomized weak coherent pulses. 

Despite these limitations, our results provide a unified framework
for analyzing source and detector correlations in QKD. We present
the results as self-contained theorem statements to facilitate
their application to concrete protocols and provide a basis for
further extensions.

\acknowledgments{We thank Jerome Wiesemann for valuable discussions and feedback on this manuscript. We thank Shlok Nahar and Norbert Lütkenhaus for interesting discussions. This work was conducted at the Institute for Quantum Computing, University of Waterloo, which is funded
by the Government of Canada through ISED. AA acknowledges support provided by  Natural Sciences and Engineering
Research Council of Canada (NSERC) under the Discovery Grants Program, Grant No. 341495.
DT was supported by the NSERC Alliance Project QUINT during his time at the Institute for Quantum Computing. DT also acknowledges support by NSERC (ALLRP-578455-2022).

All proofs and results in this work were obtained without the use of any AI tools; AI tools (primarily GPT Astra and Sol) were only used for editing and proof checking purposes.
}

    \section*{Code availability}
\label{sec:code_availability}

\noindent We will provide the code used to compute the key rates for this work in Ref.~\cite{code}, for which we use the open source OpenQKDSecurity package \cite{burniston_software_2024}. The code utilized in this work was obtained by small modifications to the code used in Ref.~\cite{wiesemann_securityframework_2026}, which was written by John Burniston and Jerome Wiesemann.

\bibliography{masterbib_QKD}
\appendix
\section{Weighted {\Renyi} chain rules}\label{app:weighted-chain-rules}
We use the following lemma in our analysis of source correlations to combine tradeoff functions on the interleaved blocks. 

\begin{lemma} \label{lem:weighted-two-block-chain-rule}
    Let $\rho\in\dop{=}(S_oS_e\CP_o\CP_eE)$ be classical on $\CP_o\CP_e$. Let $f_o,f_e$ be tradeoff functions defined on $\CP_o,\CP_e$ respectively. Take any $\alpha,\beta,\gamma\in(\frac12,1)\cup(1,\infty)$, such that 
    \begin{align}\label{eq:interpolation parameters}
        \frac{\alpha}{\alpha-1}=\frac{\beta}{\beta-1}+\frac{\gamma}{\gamma-1}.
    \end{align}
    Then, we have
    \begin{align}
        &\Halpha[f_o+f_e](S_oS_e|\CP_o\CP_eE)_\rho\geq\Halpha[f_o]<\beta>(S_o|S_e\CP_o\CP_eE)_\rho+\Halpha[f_e]<\gamma>(S_e|\CP_o\CP_eE)_\rho,\qquad\qquad\operatorname{if}\quad(\alpha-1)(\beta-1)(\gamma-1)>0,\nonumber\\
        &\Halpha[f_o+f_e](S_oS_e|\CP_o\CP_eE)_\rho\leq\Halpha[f_o]<\beta>(S_o|S_e\CP_o\CP_eE)_\rho+\Halpha[f_e]<\gamma>(S_e|\CP_o\CP_eE)_\rho,\qquad\qquad\operatorname{if}\quad(\alpha-1)(\beta-1)(\gamma-1)<0.
    \end{align}
    where $f_o + f_e$ is a tradeoff function on $\CP_o \CP_e$.
    \begin{proof}
       We prove the result for $\alpha,\beta,\gamma>1$. The other cases follow similarly by using the corresponding direction of the unweighted chain rule, replacing H\"older's inequality by its reverse form when $(\beta-1)(\gamma-1)<0$, and keeping track of the sign of $\alpha/(1-\alpha)$.

Throughout the proof, sums are restricted to outcomes $\cP_o\cP_e$ with $\rho(\cP_o\cP_e)>0$. We begin by expanding the $f$-weighted entropy on the LHS as
\begin{align}\label{eq:expansion1}
\Halpha[f_o+f_e](S_oS_e|\CP_o\CP_eE)_\rho&=\frac{\alpha}{1-\alpha}\log\left(\sum_{\cP_o\cP_e}\rho(\cP_o\cP_e)
2^{\frac{1-\alpha}{\alpha}\left(\Halpha(S_oS_e|E)_{\rho_{|\cP_o\cP_e}}-f_o(\cP_o)-f_e(\cP_e)\right)}\right)\nonumber\\
&\quad\geq\frac{\alpha}{1-\alpha}\log\left(\sum_{\cP_o\cP_e}\rho(\cP_o\cP_e)
2^{\frac{1-\alpha}{\alpha}\left(\Halpha<\beta>(S_o|S_eE)_{\rho_{|\cP_o\cP_e}}+\Halpha<\gamma>(S_e|E)_{\rho_{|\cP_o\cP_e}}-f_o(\cP_o)-f_e(\cP_e)\right)}\right).
\end{align}
The inequality follows from~\cite[Theorem~1]{Dupuis2015}, applied to each normalized conditional state $\rho_{|\cP_o\cP_e}$. Let us define vectors $\bold{A}_\beta$ and $\bold{B}_\gamma$ with components
\begin{align}\label{eq:weighted_vectors}
\bold{A}_\beta(\cP_o\cP_e)&\coloneq
2^{\frac{1-\alpha}{\alpha}\left(\Halpha<\beta>(S_o|S_eE)_{\rho_{|\cP_o\cP_e}}-f_o(\cP_o)\right)},\nonumber\\
\bold{B}_\gamma(\cP_o\cP_e)&\coloneq
2^{\frac{1-\alpha}{\alpha}\left(\Halpha<\gamma>(S_e|E)_{\rho_{|\cP_o\cP_e}}-f_e(\cP_e)\right)}.
\end{align}
Then,
\begin{align}\label{eq:vectorize}
&\sum_{\cP_o\cP_e}\rho(\cP_o\cP_e)
2^{\frac{1-\alpha}{\alpha}\left(\Halpha<\beta>(S_o|S_eE)_{\rho_{|\cP_o\cP_e}}+\Halpha<\gamma>(S_e|E)_{\rho_{|\cP_o\cP_e}}-f_o(\cP_o)-f_e(\cP_e)\right)}=\sum_{\cP_o\cP_e}\rho(\cP_o\cP_e)\bold{A}_\beta(\cP_o\cP_e)\bold{B}_\gamma(\cP_o\cP_e).
\end{align}
Set
\begin{align}
r:=\frac{\alpha/(\alpha-1)}{\beta/(\beta-1)},\qquad
s:=\frac{\alpha/(\alpha-1)}{\gamma/(\gamma-1)}.\nonumber
\end{align}
By Eq.~\eqref{eq:interpolation parameters}, we have $1/r+1/s=1$. Since $\alpha,\beta,\gamma>1$, both $r$ and $s$ are greater than one. Applying H\"older's inequality to the RHS of Eq.~\eqref{eq:vectorize} gives
\begin{align}\label{eq:weighted_holder}
&\sum_{\cP_o\cP_e}\rho(\cP_o\cP_e)\bold{A}_\beta(\cP_o\cP_e)\bold{B}_\gamma(\cP_o\cP_e)\leq
\left(\sum_{\cP_o\cP_e}\rho(\cP_o\cP_e)\bold{A}_\beta(\cP_o\cP_e)^{r}\right)^{1/r}
\left(\sum_{\cP_o\cP_e}\rho(\cP_o\cP_e)\bold{B}_\gamma(\cP_o\cP_e)^{s}\right)^{1/s}.
\end{align}
Taking logarithms and multiplying by $\alpha/(1-\alpha)<0$, we obtain
\begin{align}\label{eq:log holder}
\frac{\alpha}{1-\alpha}\log\left(\sum_{\cP_o\cP_e}\rho(\cP_o\cP_e)\bold{A}_\beta(\cP_o\cP_e)\bold{B}_\gamma(\cP_o\cP_e)\right)&\geq\frac{\beta}{1-\beta}\log\left(\sum_{\cP_o\cP_e}\rho(\cP_o\cP_e)\bold{A}_\beta(\cP_o\cP_e)^{r}\right)+\frac{\gamma}{1-\gamma}\log\left(\sum_{\cP_o\cP_e}\rho(\cP_o\cP_e)\bold{B}_\gamma(\cP_o\cP_e)^{s}\right),
\end{align}
where we used
\begin{align}
\frac{\alpha}{(1-\alpha)r}=\frac{\beta}{1-\beta},\qquad
\frac{\alpha}{(1-\alpha)s}=\frac{\gamma}{1-\gamma}.\nonumber
\end{align}
Combining Eqs.~\eqref{eq:expansion1}, \eqref{eq:vectorize}, and~\eqref{eq:log holder}, we have
\begin{align}\label{eq:weighted_chain_conclusion}
\Halpha[f_o+f_e](S_oS_e|\CP_o\CP_eE)_\rho&\geq\frac{\beta}{1-\beta}\log\left(\sum_{\cP_o\cP_e}\rho(\cP_o\cP_e)\bold{A}_\beta(\cP_o\cP_e)^{r}\right)
+\frac{\gamma}{1-\gamma}\log\left(\sum_{\cP_o\cP_e}\rho(\cP_o\cP_e)\bold{B}_\gamma(\cP_o\cP_e)^{s}\right)\nonumber\\
&\quad=\frac{\beta}{1-\beta}\log\left(\sum_{\cP_o\cP_e}\rho(\cP_o\cP_e)
2^{\frac{1-\beta}{\beta}\left(\Halpha<\beta>(S_o|S_eE)_{\rho_{|\cP_o\cP_e}}-f_o(\cP_o)\right)}\right)\nonumber\\
&\qquad\quad+\frac{\gamma}{1-\gamma}\log\left(\sum_{\cP_o\cP_e}\rho(\cP_o\cP_e)
2^{\frac{1-\gamma}{\gamma}\left(\Halpha<\gamma>(S_e|E)_{\rho_{|\cP_o\cP_e}}-f_e(\cP_e)\right)}\right)\nonumber\\
&\quad=\Halpha[f_o]<\beta>(S_o|S_e\CP_o\CP_eE)_\rho
+\Halpha[f_e]<\gamma>(S_e|\CP_o\CP_eE)_\rho.
\end{align}

    \end{proof}
\end{lemma}

\begin{corollary}\label{cor:weighted-multiblock-chain-rule}
Let $\rho$ be classical on $C_1\cdots C_L$, let $f_r$ be a tradeoff function on $C_r$, and let $S_1,\ldots,S_L$ be quantum registers. If $\alpha,\beta_1,\ldots,\beta_L>1$ satisfy
\begin{equation}
\frac{\alpha}{\alpha-1}
=\sum_{r=1}^{L}\frac{\beta_r}{\beta_r-1},
\end{equation}
then
\begin{equation}
\Halpha[\sum_{r=1}^{L}f_r]
(S_1\cdots S_L|C_1\cdots C_LE)_\rho
\geq
\sum_{r=1}^{L}
\Halpha[f_r]<\beta_r>
(S_r|S_{r+1}\cdots S_LC_1\cdots C_LE)_\rho,
\label{eq:weighted-multiblock-chain-rule}
\end{equation}
where $\sum_{r=1}^{L}f_r$ is a tradeoff function on $C_1 \dots C_L$. 
\end{corollary}

\begin{proof}
The proof follows from a straightforward inductive argument and \cref{lem:weighted-two-block-chain-rule}.
\end{proof}

\section{Computing the effective marginal states}
\label{app:keyrates}
In this appendix, we will obtain the set of marginal states that will appear in \cref{eq:fhatfulldefinedsourcedetcombined}, over all possible values of measurement outcomes of the measured rounds $u_{\otherinterleavedrounds{r}}$, for the correlated source model described in \cref{sec:plots}. For simplicity, we ignore the technical issues arising from blocks at the start and end of the protocol. We will also have to modify our notations slightly.

Fix a selected round $j = \selectedround{k}{r}$, and condition on the
setting choices $u=u_{\mathcal I_{\neq r}}$. For the settings in the window $\max\{1,j-l_{\mathrm c}^{\mathrm{src}}\},\ldots,j+\lcsrc$,
define
\begin{equation}
    i_q(x,u)\coloneqq
    \begin{cases}
        x, & q=j,\\
        u_q, & q\ne j.
    \end{cases}
\end{equation}
Using the signal states defined in the source model, the actual joint state emitted in rounds $j,\ldots,j+\lcsrc$,
conditioned on $i_j=x$ and the other settings $u$, is
\begin{equation}
    \ket{\Phi_x^{\mathrm{act}}(u)}_{\blockversion{A}'_{k,r}}
    \coloneqq
    \bigotimes_{q=j}^{j+\lcsrc}
    \ket{
        \Psi_{i_q(x,u)}^{
            (q,i_{q-l_{\mathrm c}^{\mathrm{src}}}^{q-1}(x,u))
        }
    }_{A'_q}.
    \label{eq:cbb84-actual-state}
\end{equation}
For a fixed reference setting $x_*\in\{0,1,2,3\}$, define
\begin{equation}
    \ket{\Phi_x^{\mathrm{ref}}(u)}_{\blockversion{A}'_{k,r}}
    \coloneqq
    \ket{
        \Psi_x^{(j,u_{j-l_{\mathrm c}^{\mathrm{src}}}^{j-1})}
    }_{A'_j}
    \otimes
    \bigotimes_{q=j+1}^{j+\lcsrc}
    \ket{
        \Psi_{u_q}^{
            (q,i_{q-l_{\mathrm c}^{\mathrm{src}}}^{q-1}(x_*,u))
        }
    }_{A'_q}.
    \label{eq:cbb84-reference-state}
\end{equation}
Notice that the reference states are such that only the state on $A'_j$ depends on the setting choice $i_j=x$, and thus the overlaps $\braket{  \ket{\Phi_x^{\mathrm{ref}}(u)}}{  \ket{\Phi_{x'}^{\mathrm{ref}}(u)}}$ corresponds to the overlap of ideal BB84 states. 

By the overlap calculation in Appendix~C, Eq.~(26), of
Ref.~\cite{curraslorenzo_phaseerrorcorrelated_2026}, for every
current setting $x$ and conditioned setting string $u$,
\begin{align}
    \left|
        \left\langle
            \Phi_x^{\mathrm{ref}}(u)
            \middle|
            \Phi_x^{\mathrm{act}}(u)
        \right\rangle
    \right|^2
    &=
    \prod_{l=1}^{\lcsrc}
    \cos^2\!\bigl(\delta_l(x)-\delta_l(x_*)\bigr)
    \ge 1-\varepsilon_{\mathrm{src}},
    \label{eq:cbb84-fidelity}\\
    \varepsilon_{\mathrm{src}}
    &\coloneqq
    \min\!\left\{
        1,\sum_{l=1}^{l_{\mathrm c}^{\mathrm{src}}}\xi_l
    \right\}
    =
    \min\!\left\{
        1,\,
        \xi_1
        \frac{
            1-e^{-C_{\mathrm{decay}}l_{\mathrm c}^{\mathrm{src}}}
        }{
            1-e^{-C_{\mathrm{decay}}}
        }
    \right\},
    \label{eq:cbb84-infidelity}
\end{align}
where the geometric expression assumes $C_{\mathrm{decay}}>0$. 

Using this, it is straightforward to verify that the set of possible marginals in \cref{lemma:theblockstates} on the $A_j$ system is contained in the set of possible marginals when the emitted state is close in fidelity to the ideal BB84 state, where the fidelity is lower bounded by $1-\varepsilon_\mathrm{src}$.

\end{document}